\documentclass[11pt]{article}

\usepackage{fullpage,times,amssymb}
\newtheorem{definition}{Definition} 
\newtheorem{fact}{Fact}    
\newtheorem{theorem}{Theorem}    
\newtheorem{result}{Result}    
\newtheorem{corollary}{Corollary}    
\newtheorem{lemma}{Lemma}      
\newtheorem{claim}{Claim}      
\newtheorem{proposition}{Proposition}      
\newcommand{\qed}{\hfill{$\rule{6pt}{6pt}$}} 
\newenvironment{proof}{\noindent{\bf Proof}:}{\qed}

\newcommand{\defeq}{:=}

\newcommand{\ket}[1]{| #1 \rangle}
\newcommand{\bra}[1]{\langle #1 |}
\newcommand{\ketbra}[1]{| #1 \rangle \langle #1 |}
\newcommand{\unibraket}[2]{\langle #1 | #2 | #1 \rangle}
\newcommand{\braket}[2]{\langle #1 | #2 \rangle}

\newcommand{\E}{\mbox{{\rm E} }}

\newcommand{\Tr}{\mbox{{\rm Tr} }}
\newcommand{\parttr}[1]{\mbox{${\rm Tr}_{#1}$ }}

\newcommand{\trnorm}[1]{\left\| #1 \right\|_{\mathrm{tr}}}
\newcommand{\totvar}[1]{\left\| #1 \right\|_1}

\newcommand{\cF}{{\cal F}}
\newcommand{\cH}{{\cal H}}
\newcommand{\cK}{{\cal K}}
\newcommand{\cL}{{\cal L}}
\newcommand{\cP}{{\cal P}}
\newcommand{\cT}{{\cal T}}
\newcommand{\cX}{{\cal X}}
\newcommand{\cY}{{\cal Y}}
\newcommand{\cZ}{{\cal Z}}

\newcommand{\naturals}{\mathbb N}
\newcommand{\reals}{\mathbb R}
\newcommand{\complexes}{\mathbb C}

\newcommand{\event}{{\cal E}}

\newcommand{\identity}{\leavevmode\hbox{\small1\kern-3.8pt\normalsize1}}

\newcommand{\supp}{{\rm supp}}

\newcommand{\vc}{{\rm VC}}

\newcommand{\threearray}[6]{
	\left[
	\begin{array}{c c r}
		\parbox{1.5cm}
                {\LARGE \ \ $#1$\ } & #2      & #3 \\ \\
		#2^\dag             & #4      & #5 \\
		#3^\dag             & #5^\dag & #6
	\end{array}
	\right]}

\newcommand{\bigtwoarray}[3]{
	\left[
	\begin{array}{c c r}
		\parbox{1.5cm}
                {\raisebox{-1cm}{\Huge \ \ \ $#1$}} & & #2 \\ \\
		#2^\dag                             & & #3
	\end{array}
	\right]}

\newcommand{\twoarray}[3]{
	\left[
	\begin{array}{c c}
	        #1      & #2 \\
		#2^\dag & #3
	\end{array}
	\right]}

\title{A theorem about relative entropy of quantum states with an
application to privacy in quantum communication}
\author{
Rahul Jain \thanks{Email: {\sf
rjain@cs.uwaterloo.ca}. Most of this work was done while the author was at
Tata Institute of Fundamental Research, Mumbai, India, and partly at
U.C. Berkeley, California, USA.} \\ 
Institute for Quantum Computing \\
University of Waterloo \\
Waterloo, Canada.
\and
Jaikumar Radhakrishnan  \thanks{Email: {\sf
jaikumar@tifr.res.in} } \\ 
School of Technology and Computer Science \\
Tata Institute of Fundamental Research \\
Mumbai, India
\and
Pranab Sen  \thanks{Email:  {\sf
pgdsen@tcs.tifr.res.in}. Most of this work was done while the author was 
at Laboratoire de Recherche en Informatique, Universit\'e de Paris--Sud, 
Orsay, France.
}\\ 
School of Technology and Computer Science \\
Tata Institute of Fundamental Research \\
Mumbai, India
}
\date{}

\begin{document}
\maketitle

\begin{abstract}
\noindent We prove the following theorem about relative entropy of quantum
states.
\begin{quote}
{\em Substate theorem:} Let $\rho$ and $\sigma$ be quantum states in
the same Hilbert space with
relative entropy $S(\rho \| \sigma) \defeq \Tr \rho (\log \rho -\log
\sigma) = c$. Then for all $\epsilon>0$, there is a state $\rho'$ such
that the trace distance 
$\trnorm{\rho' - \rho} \defeq \Tr \sqrt{(\rho' - \rho)^2} \leq \epsilon$,
and $\rho'/2^{O(c/\epsilon^2)} \leq \sigma$.
\end{quote}
It states that if the relative entropy of $\rho$ and $\sigma$ is
small, then there is a state $\rho'$ close to $\rho$, i.e. with small
trace distance $\trnorm{\rho' - \rho}$, that when scaled down by a factor
$2^{O(c)}$ `sits inside', or becomes a `substate' of, $\sigma$.  This
result has several applications in quantum communication complexity and
cryptography. Using the substate theorem, we derive a privacy trade-off
for the {\em set membership problem} in the two-party quantum
communication model.  Here Alice is given a subset $A \subseteq [n]$,
Bob an input $i \in [n]$, and they need to determine if $i \in A$.
\begin{quote}
{\em Privacy trade-off for set membership:} In any
two-party quantum communication protocol for the set membership
problem, if Bob reveals only $k$ bits of information about his input,
then Alice must reveal at least $n/2^{O(k)}$ bits of information about
her input.
\end{quote} 
We also discuss relationships between various information
theoretic quantities that arise naturally in the context of the
substate theorem.
\end{abstract}

\section{Introduction}
The main contribution of this paper is a theorem, called the substate
theorem; it states, roughly, that if the relative entropy, $S(\rho \|
\sigma) \defeq \Tr \rho (\log \rho -\log \sigma)$, of two quantum states 
$\rho$ and $\sigma$ is at most $c$,
then there a state $\rho'$ close to sigma such that $\rho'/2^{O(c)}$
{\em sits inside} $\sigma$. This implies that, as we will formalise later,
state $\sigma$ can `masquerade' as state $\rho$ with probability
$2^{-O(c)}$ in many situations. Before we
discuss the substate theorem, let us first see a setting in which it is
applied in order to get some motivation. This application concerns 
the trade-off in privacy in
two-party quantum communication protocols for the set membership
problem~\cite{miltersen:roundelim}. After that, we discuss the 
substate theorem
proper followed by a brief description of several subsequent applications
of the theorem.

\subsection{The set membership problem}
\newcommand{\set}{\mathsf{SetMemb}}
\begin{definition}
In the {\em set membership problem} $\set_n$, Alice is given a subset $A
\subseteq [n]$ and Bob an element $i \in [n]$. The two parties are 
required to exchange messages according to a fixed protocol in order
for the last recipient of a message to determine if $i \in [n]$. We 
often think of Alice's input as a
string $x \in \{0,1\}^n$ which we view as the characteristic vector of
the set $A$; the protocol 
requires that in the end the last recipient output $x_i$. In this 
viewpoint, Bob's input $i$ is called an {\em index} and the set
membership problem is called the {\em index function problem}.
\end{definition}
The set membership problem is a fundamental problem in communication
complexity. In the classical setting, it was studied by Miltersen,
Nisan, Safra and Wigderson~\cite{miltersen:roundelim}, who showed that
if Bob sends a total of at most $b$ bits, then Alice must send
$n/2^{O(b)}$ bits. Note that this is optimal up to constants, as there
is a trivial protocol where Bob sends the first $b$ bits of his index
to Alice, and Alice replies by sending the corresponding part of her
bit string.  The proof of Miltersen {\em et al.} relied on the {\em
richness technique} they developed to analyse such protocols. However,
here is a simple round-elimination argument that gives this lower
bound, and as we will see below, this argument generalises to the
quantum setting.  Fix a protocol where Bob sends a total of at most
$b$ bits, perhaps spread over several rounds. We can assume without
loss of generality that Bob is the last recipient of a message, otherwise
we can augment the protocol by making Alice send the answer to Bob at
the end which increases Alice's communication cost by one bit.
Modify this protocol as
follows. In the new protocol, Alice and Bob use shared randomness to
guess all the messages of Bob. Alice sends her responses based on this
guess. After this, if Bob finds that the guessed messages are exactly
what he wanted to send anyway, he accepts the answer given by the
original protocol; otherwise, he aborts the protocol. Thus, if the
original protocol was correct with probability $p$, the new one-round
protocol, when it does not abort, which happens with probability at
least $2^{-b}$, is correct with probability at least $p$. A
standard information theoretic argument of Gavinsky, Kempe, Regev
and de~Wolf~\cite{deWolfKempe} now shows
that in any such protocol, Alice must send $2^{-b} \cdot n (1-H(p))$ bits.

In the quantum setting, a special case of the set membership problem
was studied by Ambainis, Nayak, Ta-Shma and Vazirani~\cite{ANTV:index},
where Bob is not allowed to send any message and there is no prior
entanglement between Alice and Bob.
They referred to this as {\em quantum random access codes}, because in
this setting the problem can be thought of as Alice encoding $n$ classical
bits $x$ using qubits in such a way that Bob is able to determine any
one $x_i$ with probability at least $p \geq \frac{1}{2}$. Note that in
the quantum setting, unlike in its classical counterpart, it is
conceivable that the measurement needed to determine $x_i$ makes the
state unsuitable for determining any of the other bits $x_j$. In fact,
Ambainis {\em et al.} exhibit a quantum random access code encoding
two classical bits $(x_1, x_2)$ into one qubit such that any single
bit $x_i$ can be recovered with probability strictly greater than $1/2$,
which is impossible classically.
Their main result, however, was that any such quantum code
must have $n(1-H(p))$ qubits. They also gave a classical 
code with encoding length $n(1-H(p)) + O(\log n)$, thus showing that 
quantum random access codes provide no substantial improvement over 
classical random access codes.
     
    In this paper, we study the general set membership problem,
where Alice and Bob are allowed to exchange quantum messages over several
rounds as well as share prior entanglement.
Ashwin Nayak (private communication) observed that the
classical round elimination argument described above is applicable in
the quantum setting: if Alice and Bob share prior entanglement in the
form of EPR
pairs, then using quantum teleportation~\cite{bennett:teleportation},
Bob's messages can be assumed to be classical. Now, Alice can guess
Bob's messages, and we can combine the classical round elimination
argument above with the results on random access codes to show that
Alice must send at least $2^{-(2 b + 1)} \cdot n (1-H(p))$ qubits to Bob.

   We strengthen these results and show that this trade-off between the
communication required of Alice and Bob is in fact a trade-off in their
privacy: if a protocol has the property that Bob `leaks' only a
small number of bits of {\em information} about his input, then in that
protocol Alice must leak a large amount of information about her
input; in particular, she must send a large number of
qubits. Before we present our result, let us explain what we mean when
we say that Bob leaks only a small number of bits of information about
his input. Fix a protocol for set membership.  Assume that
Bob's input $J$ is a random element of $[n]$. Suppose Bob operates
faithfully according to the protocol, but Alice deviates from it and
manages to get her registers, say $A$, entangled with $J$: we say that
Bob leaks only $b$ bits of information about his input if the mutual
information between $J$ and $A$, $I(J : A)$, is at most $b$. This must
hold for all strategies adopted by Alice. Note that we do not assume
that Bob's messages contain only $b$ qubits, they can be arbitrarily
long. In the quantum setting, Alice has a big bag of tricks she can
use in order to extract information from Bob.  See
Section~\ref{subsec:privacy} for an example of a cheating strategy for
Alice, that exploits Alice's ability to perform quantum operations.
We show the following result.
\begin{result}[informal statement]
\label{res:privacy}
If there is a quantum protocol for the set membership problem where
Bob leaks only $b$ bits of information about his input $J$, then Alice
must leak $\Omega(n/2^{O(b)})$ bits of information about her input
$x$.  In particular, this implies that Alice must send $n/2^{O(b)}$
qubits.
\end{result}

\paragraph{Related work:}
One can compare this with work on private information
retrieval~\cite{chor:pir}.  There, one requires that the party holding
the database $x$ know nothing about the index $i$. 
Nayak~\cite{nayak:index} sketched an argument
showing that in both classical and quantum settings, the party
holding the database
has to send $\Omega(n)$ bits/qubits to the party holding the index.
Result~\ref{res:privacy}
generalises Nayak's argument and shows a trade-off between the loss in
privacy for the database user Bob, and the loss in privacy for the
database server Alice.

  Recently, Klauck~\cite{klauck:privacy} studied privacy in quantum
protocols.  In Klauck's setting, two players collaborate to compute a
function, but at any point, one of the players might decide to
terminate the protocol and try to infer something about the input of
the other player using the bits in his possession.  The players are
{\em honest but curious}: in a sense, they don't deviate from the
protocol in any way other than, perhaps, by stopping early.  In this
model, Klauck shows that there is a protocol for the {\em set
disjointness} function where neither player reveals more than $O((\log
n)^2)$ bits of information about his input, whereas in every classical
protocol, at least one of the players leaks $\Omega(\sqrt{n}/\log n)$
bits of information about his input.
Our model of privacy is more stringent. We allow malicious players who
can deviate arbitrarily from the protocol. An immediate corollary of
our result is that for the set membership problem, one of the players
must leak $\Omega(\log n)$ bits of information. This implies a similar
loss in privacy for several other problems, including the set
disjointness problem.

\paragraph{Privacy trade-off and the substate theorem:} 
We now briefly motivate the need for the substate theorem in showing
the privacy trade-off in Result~\ref{res:privacy} above. 
We know from the communication trade-off argument for set membership 
presented above that in any protocol for the problem, if Bob sends only
$b$ qubits, then Alice must send $n/2^{O(b)}$ qubits.  Unfortunately,
this argument is not applicable when the protocol does not promise
that Bob sends only $b$ qubits, but only ensures that the number of
bits of information Bob leaks is at most $b$. So, the assumption is
weaker. On the other hand, the conclusion now is stronger, for it
asserts that Alice must leak $n/2^{O(b)}$ bits of information, which
implies that she must send at least these many qubits. The above
argument relied on the fact that Alice could generate a distribution
on messages, so that every potential message of Bob is
well-represented in this distribution: if Bob's messages are classical
and $b$ bits long, the uniform distribution is such a
distribution---each $b$ bit message appears in it with probability
$2^{-b}$.  Note that we are not assuming that messages of Bob have at most
$b$ qubits, so Alice cannot guess these messages in this manner.
Nevertheless, using only the assumption that Bob leaks at most $b$
bits of information about his input, the substate theorem provides us
an alternative for the uniform distribution.  It allows us to prove
the existence of a single quantum state that Alice and Bob can 
generate without access to Bob's
input, after which if Bob is provided the input $i$, he can obtain the
correct final state with probability at least $2^{-O(b)}$ or abort if
he cannot. After this, a quantum information theoretic argument of
Gavinsky, Kempe, Regev and de Wolf~\cite{deWolfKempe} implies that Alice
must leak at least $n/2^{O(b)}$ bits of information about her input. 
The proof is discussed in detail in Section~\ref{sec:tradeoffs}.

\subsection{The substate theorem}
\label{subsec:introsubstate}
It will be helpful to first consider the classical analogue of the
substate theorem. Let $P$ and $Q$ be probability distributions on the
set $[n]$ such that their relative entropy is bounded by $c$, that is
\begin{equation}
\label{eq:defS}
 S(P \| Q) \defeq \sum_{i\in [n]} P(i) \log_2 \frac{P(i)}{Q(i)}
~\leq~c 
\end{equation}
When $c$ is small, this implies that $P$ and $Q$ are close to each other
in {\em total variation distance}; indeed, one can show that
(see e.g.~\cite[Lemma~12.6.1]{cover:infotheory})
\begin{equation}
\label{eq:SversusTr} 
\totvar{P - Q}  \defeq \sum_{i\in [n]} |P(i)-Q(i)| ~\leq~ 
\sqrt{(2 \ln 2)c}. 
\end{equation}
That is, the probability of an event $\event
\subseteq [n]$ in $P$ is close to its probability in $Q$:
$|P(\event)-Q(\event)| \leq \sqrt{(c\ln 2)/2} $.  Now consider the
situation when $c \gg 1$. In that case, expression~(\ref{eq:SversusTr}) 
becomes
weak, and it is not hard to construct examples where $\totvar{P - Q}$
is very close to $2$. Thus by bounding $\totvar{P - Q}$ alone, we
cannot infer that an event $\event$ with probability $3/4$ in
$P$ has any non-zero probability in $Q$. But is it true that when $S(P
\| Q) < +\infty$ and $P(\event) > 0$, then $Q( \event) > 0$? Yes! To
see this, let us reinterpret the expression in (\ref{eq:defS}) as the
expectation of $\log P(i)/Q(i)$ as $i$ is chosen according to
$P$. Thus, one is lead to believe that if $S(P\|Q)
\leq c < +\infty$, 
then $\log P(i)/Q(i)$ is typically bounded by $c$, that is,
$P(i)/Q(i)$ is typically
bounded by $2^c$. One can formalise this intuition and
show, for all $r \geq 1$,
\begin{equation}
\label{eq:weaksubstate}
\Pr_{i\in P}\left[\frac{P(i)}{Q(i)} > 2^{r(c+1)}\right] < \frac{1}{r}.
\end{equation}
We now briefly sketch a proof of the above inequality.
Let ${\sf Good} \defeq \{i: P(i)/2^{r(c+1)} \leq Q(i)\}$,
${\sf Bad} \defeq [n] \setminus {\sf Good}$. By concavity of
the logarithm function, we get 
\[
P({\sf Good}) \log \frac{P({\sf Good})}{Q({\sf Good})} +
P({\sf Bad}) \log \frac{P({\sf Bad})}{Q({\sf Bad})} 
\leq S(P \| Q) \leq c.
\]
By elementary calculus, 
$P({\sf Good}) \log \frac{P({\sf Good})}{Q({\sf Good})} > -1$. 
Thus we get $P({\sf Bad}) \cdot r (c+1) < c+1$, proving the 
above inequality.

We now define a new probability distribution $P'$ as follows:
\[
P'(i) \defeq
\left\{
\begin{array}{l l}
\frac{P(i)}{P({\sf Good})} & i \in {\sf Good} \\
0                          & i \in {\sf Bad}
\end{array}
\right.,
\]
that is, in $P'$ we just discard the bad values of $i$
and renormalise. Now, $\frac{r-1}{r 2^{r(c+1)}}P'$ is dominated by $Q$
everywhere.  We have thus shown the classical analogue of the
desired substate theorem.
\paragraph{Result~\ref{res:substate}' (Classical substate theorem)}
{\em
Let $P, Q$ be probability distributions on the same sample space with
$S(P \| Q)\leq c$. Then for all $r > 1$, there exist distributions
$P', P''$  such that $\totvar{P-P'} \leq \frac{2}{r}$ and
$Q = \alpha P' + (1-\alpha) P''$,
where $\alpha \defeq \frac{r-1}{r 2^{r(c+1)}}$.
}

Let us return to our event $\event$ that occurred with some small
probability $p$ in $P$. Now, if we take $r$ to be $2/p$, then $\event$
occurs with probability at least $p/2$ in $P'$, and hence appears with
probability $p/2^{O(c/p)}$ in $Q$. Thus, we have shown that even though
$P$ and $Q$ are far apart as distributions, events that have positive
probability, no matter how small, in $P$, continue to have positive
probability in $Q$. 

The main contribution of this paper is a quantum analogue of
Result~\ref{res:substate}'. To state it, we recall that the
relative entropy of two quantum states $\rho, \sigma$ in the same
Hilbert space is defined as
$S(\rho \| \sigma) \defeq \Tr \rho (\log \rho - \log \sigma)$, and
the {\em trace distance} between them is defined as
$\trnorm{\rho -\rho'} \defeq \Tr \sqrt{(\rho - \rho')^2}$.
\begin{result}[Quantum substate theorem]
\label{res:substate}
Suppose $\rho$ and $\sigma$ are
quantum states in the same Hilbert space with $S(\rho \| \sigma) \leq c$. 
Then for all $r >1$, there
exist states $\rho', \rho''$ such that 
$\trnorm{\rho -\rho'} \leq \frac{2}{\sqrt{r}}$ and 
$\sigma = \alpha \rho' + (1-\alpha)\rho''$,
where $\alpha \defeq \frac{r-1}{r 2^{rc'}}$ and
$c' \defeq c + 4 \sqrt{c+2} + 2 \log (c+2) + 5$.
\end{result}
The quantum substate theorem has been stated above in a form that brings 
out the analogy with
the classical statement in Result~\ref{res:substate}'. In 
Section~\ref{sec:substate}, we have
a more nuanced statement which is often better suited for applications.

\paragraph{Remark:} 
Using the quantum substate theorem and arguing as above, one can conclude 
that if an event $\event$ has probability $p$ in $\rho$, then its
probability $q$ in $\sigma$ is at least $q \geq \frac{p}{2^{O(c/p^2)}}$,
$c = S(\rho \| \sigma)$. Actually, one can show the stronger result
that $q \geq \frac{p}{2^{O(c/p)}}$ as follows. 
Using the fact that relative 
entropy cannot increase after doing a measurement, we get 
\[
p \log \frac{p}{q} + (1-p) \log \frac{1-p}{1-q} \leq S(\rho \| \sigma)
\leq c. 
\]
We now argue as in the proof of Result~\ref{res:substate}' to show the
stronger lower bound on $q$. 

In view of this, one may wonder if there is any motivation at all in
proving a quantum substate theorem. Recall
however, that the quantum substate theorem 
gives a structural relationship between $\rho$ and $\sigma$ which is useful
in many applications e.g. privacy trade-off for set membership discussed
earlier. It does not seem possible in these applications to replace
this structural relationship by considerations about the relative 
probabilities of an event $\event$ in $\rho$ and $\sigma$. In our
privacy trade-off application, $\sigma$ plays the role of the 
state that Alice and Bob can generate without access to Bob's input, 
and $\rho$ plays the role of the correct final state of Bob in the
protocol. To prove the trade-off, $\sigma$ should be able to `masquerade'
as $\rho$ with probability $2^{-O(b)}$, $b$ being the amount of 
information Bob leaks about his input. Also, Bob should {\em know} whether
the `masquerade' succeeded or not so that he can abort if it fails, and
it is this requirement that needs the substate property.

\bigskip

The ideas used to arrive at Result~\ref{res:substate}' do not
immediately generalise to prove Result~\ref{res:substate}, because 
$\rho$ and
$\sigma$ need not be simultaneously diagonalisable.  As it turns out,
our proof of the quantum substate theorem takes an indirect route. 
First, by
exploiting the Fuchs and Caves~\cite{fuchs:fidelity} characterisation
of fidelity and a minimax theorem of game theory, we obtain a
`lifting' theorem about an `observational' version of relative
entropy; this statement is interesting on its own.  Using this
`lifting' theorem, and a connection between the `observational'
version of relative entropy and actual relative entropy, we argue that
it is enough to verify the original statement when $\rho$ and $\sigma$
reside in a two-dimensional space and $\rho$ is a pure state. The two
dimensional case is then established by a direct computation.

\subsection{Other applications of the substate theorem}
The conference version of this paper~\cite{jain:substate}, in which
the substate theorem was first announced, described two
applications of the theorem.
The first application provided tight privacy trade-offs
for the set membership problem, which we have discussed above. This
application is a good illustration of the use of the substate theorem,
for several applications have the same structure. The second
application showed tight lower bounds for the {\em pointer chasing
problem}~\cite{NisanWigderson:ptr, KNTZ:ptr}, thereby establishing that the
lower bounds shown by Ponzio, Radhakrishnan and Venkatesh~\cite{ponzio:ptr}
in the
classical setting are valid also for quantum protocols without prior
entanglement.
   
Subsequent to \cite{jain:substate}, several applications of the 
classical and quantum substate theorems 
have been discovered. We briefly describe these results
now. Earlier, in related but independent work Chakrabarti, Shi, Wirth and
Yao~\cite{chakrabarti:icost} 
discovered their very influential {\em information cost}
approach for obtaining {\em direct sum} results in communication
complexity. Jain, Radhakrishnan and Sen~\cite{jain:directsum} 
observed that the
arguments used by Chakrabarti {\em et al.} could be derived more
systematically using the classical substate theorem; this approach
allowed them to extend Chakrabarti {\em et al.}'s direct sum results, 
which applied only to one-round
and simultaneous message protocols under product distributions on inputs, 
to two-party
multiple round protocols under product distributions on inputs. Ideas
from \cite{jain:directsum} were then applied by Chakrabarti and 
Regev~\cite{chakrabarti:ann} to
obtain their tight lower bound on data structures for the {\em approximate
nearest neighbour problem} on the Hamming cube. 

The quantum substate theorem, the main result of this
paper, has also found several other applications.  Jain, Radhakrishnan
and Sen~\cite{jain:entangred} used it to show how any two-party 
multiple round quantum protocol where Alice leaks only $a$
bits of information about her input and Bob leaks only $b$ bits of
information about his, can be transformed to a one-round quantum protocol
with prior entanglement
where Alice transmits just $a2^{O(b)}$ bits to Bob. 
Note that plain Schumacher compression~\cite{schumacher}
cannot be used to prove such a
result, since we require a `one-shot' as opposed to an asymptotic result,
there can be interaction in a general communication protocol, as well
as the case that the reduced state of any single party can be mixed. 
Jain {\em et al.}'s compression result gives an alternative
proof of Result~\ref{res:privacy}, because the work of 
Ambainis {\em et al.}~\cite{ANTV:index} implies that 
in any such protocol for set
membership Alice must send $\Omega(n)$ bits to Bob. 
Jain {\em et al.} also used the classical and quantum substate
theorems to prove worst case direct sum results for simultaneous message 
and one round classical and quantum protocols, improving on 
\cite{jain:directsum}.
More recently, using the quantum substate theorem
Jain~\cite{jain:remote} obtained a nearly tight characterisation of
the communication complexity of {\em remote state preparation}, an area
that has received considerable attention lately.  The substate theorem
has also found application in the study of quantum cryptographic
protocols: using it, Jain~\cite{jain:quantumstring}
showed nearly tight bounds on the {\em binding-concealing} trade-offs
for {\em quantum string commitment} schemes. 

\subsection{Organisation of the rest of the paper}
In the next section, we
recall some basic facts from classical and quantum information theory
that will be used in the rest of the paper.
In Section~\ref{sec:tradeoffs}, we
formally define our model of privacy loss in quantum communication
protocols and prove our privacy trade-off result for set membership
assuming the substate theorem.  In
Section~\ref{sec:substate}, we give the actual statement of the
substate theorem that is used in our privacy trade-offs, and a
complete proof for it. Sections~\ref{sec:tradeoffs} and \ref{sec:substate}
may be read independently of each other.
In Section~\ref{sec:conclusion} we mention some open
problems, and finally in the appendix we discuss relationships between
various information theoretic quantities that arise naturally in the
context of the substate theorem. The appendix may be read independently
of Section~\ref{sec:tradeoffs}.

\section{Information theory background}
We now recall some basic definitions and facts from classical
and quantum information
theory, which will be useful later.
For excellent introductions to classical and quantum information
theory, see the books by Cover and Thomas~\cite{cover:infotheory}
and Nielsen and Chuang~\cite{nielsen:quant} respectively.

In this paper, all functions will have finite domains and ranges,
all sample spaces will be finite, all random variables
will have finite range and all Hilbert spaces
finite dimensional. All logarithms are taken to base two.
We start off by recalling the definition of a quantum state.
\begin{definition}[Quantum state]
A quantum state or a density matrix in a Hilbert space $\cH$ is a
Hermitian, positive semidefinite operator on $\cH$ with unit trace.
\end{definition}
Note that a classical probability distribution can be thought of as a
special case of a quantum state with diagonal density matrix.
An important class of quantum states are what are known as {\em pure} 
states, which are states 
of the form $\ketbra{\psi}$, where $\ket{\psi}$ is a unit vector in
$\cH$. Often, we abuse notation and refer to $\ket{\psi}$ itself
as the pure quantum state; note that this notation is ambiguous up to
a multiplicative unit complex number. 

Let $\cH, \cK$ be two Hilbert spaces and $\omega$ a quantum state in
the {\em bipartite system} $\cH \otimes \cK$. The {\em reduced} quantum
state of $\cH$ is given by {\em tracing out} $\cK$, also
known as the {\em partial trace} 
$
\parttr{\cK} \omega \defeq 
\sum_k (\identity_{\cH} \otimes \bra{k}) \omega
       (\identity_{\cH} \otimes \ket{k})
$
where $\identity_{\cH}$ is the identity operator on $\cH$ and
the summation is over an orthonormal basis for $\cK$. It is
easy to see that the partial trace is independent of the choice of
the orthonormal basis for $\cK$. For a quantum state $\rho$ in $\cH$,
any quantum state $\omega$ in $\cH \otimes \cK$
such that $\parttr{\cK} \omega = \rho$ is said to be an {\em extension} 
of $\rho$ in $\cH \otimes \cK$; if $\omega$ is pure, it is said, more
specifically, to be a {\em purification}.

We next define a POVM element, which formalises the notion of a single
outcome of a general measurement on a quantum state.
\begin{definition}[POVM element]
A POVM (positive operator valued measure) element
$F$ on Hilbert space $\cH$ is a Hermitian positive semidefinite 
operator on $\cH$ such that $F \leq \identity$, where $\identity$ is 
the identity operator on $\cH$.
\end{definition}
If $\rho$ is a quantum state in $\cH$, the
success probability of $\rho$ under POVM element $F$ is given by
$\Tr (F \rho)$.

We now define a POVM which represents the most general form of a 
measurement allowed by quantum mechanics.
\begin{definition}[POVM]
A POVM $\cF$ on Hilbert space $\cH$ is a finite set of POVM 
elements $\{F_1, \ldots, F_k\}$ on
$\cH$ such that $\sum_{i=1}^k F_i = \identity$, where $\identity$ is the
identity operator on $\cH$.
\end{definition}
If $\rho$ is a quantum state in $\cH$, 
let $\cF \rho$ denote the probability distribution
$\{p_1, \ldots, p_k\}$ on $[k]$, where $p_i \defeq \Tr (F_i \rho)$.

Typically, the distance between two probability distributions $P, Q$ on the
same sample space $\Omega$ is measured in terms of the {\em total
variation distance} defined as 
$\totvar{P - Q} \defeq \sum_{i \in \Omega} |P(i) - Q(i)|$. 
The quantum analogue of the total variation distance is
known as the {\em trace distance}.
\begin{definition}[Trace distance]
Let $\rho, \sigma$ be quantum states in the same Hilbert space. Their
trace distance is defined as
$\trnorm{\rho - \sigma} \defeq \Tr \sqrt{(\rho - \sigma)^2}$.
\end{definition}
If we think of probability distributions as diagonal density matrices,
then the trace distance between them is nothing but their total variation
distance. For pure states $\ket{\psi}, \ket{\phi}$ it is easy to
see that their trace distance is given by
$
\trnorm{\ketbra{\psi} - \ketbra{\phi}} = 
 2 \sqrt{1 - |\braket{\psi}{\phi}|^2}.
$
The following fundamental fact shows that the
trace distance between two density matrices
bounds how well one can distinguish
between them by a POVM. A proof can be found in \cite{aharonov:mixed}.
\begin{fact}
\label{fact:totvartrace}
Let $\rho, \sigma$ be density matrices in the same 
Hilbert space $\cH$. Let $\cF$ be a POVM on $\cH$. Then,
$\totvar{\cF \rho - \cF \sigma} \leq \trnorm{\rho - \sigma}$.
Also, there is a two-outcome orthogonal measurement that achieves
equality above.
\end{fact}

Another measure of distinguishability between two probability distributions
$P, Q$ on the same sample space $\Omega$ is the 
{\em Bhattacharya distinguishability coefficient} defined as
$B(P, Q) \defeq \sum_{i \in \Omega} \sqrt{P(i) Q(i)}$. Its quantum
analogue is known as {\em fidelity}. We will need several facts about
fidelity in order to prove the quantum substate theorem.
\begin{definition}[Fidelity]
Let $\rho$, $\sigma$ be density matrices in the same 
Hilbert space $\cH$. Their fidelity is defined as
$
B(\rho, \sigma) \defeq \Tr \sqrt{\sqrt{\rho} \sigma \sqrt{\rho}}.
$
\end{definition}
The fidelity, or sometimes its square, is also referred to as the
``transition probability'' of Uhlmann.
For probability distributions, the fidelity turns out to be the same
as their Bhattacharya distinguishability coefficient. 
Jozsa~\cite{jozsa:fidelity}
gave an elementary proof for finite dimensional Hilbert
spaces of the following basic and remarkable property about fidelity.
\begin{fact}
\label{fact:jozsa}
Let $\rho, \sigma$ be density matrices in the same 
Hilbert space $\cH$. Then,
$
B(\rho, \sigma)=\sup_{\cK, \ket{\psi}, \ket{\phi}} |\braket{\psi}{\phi}|,
$
where $\cK$ ranges over all Hilbert spaces and
$\ket{\psi}, \ket{\phi}$ range over all purifications of
$\rho, \sigma$ respectively in $\cH \otimes \cK$. Also, for any 
Hilbert space $\cK$ such that $\dim (\cK) \geq \dim (\cH)$, there
exist purifications
$\ket{\psi}, \ket{\phi}$ of $\rho, \sigma$ in $\cH \otimes \cK$,
such that $B(\rho, \sigma) = |\braket{\psi}{\phi}|$.
\end{fact}
We will also need the following fact about fidelity, proved
by Fuchs and Caves~\cite{fuchs:fidelity}. 
\begin{fact}
\label{fact:fuchscaves}
Let $\rho, \sigma$ be density matrices in the same 
Hilbert space $\cH$. Then
$
B(\rho, \sigma) = \inf_{\cF} B(\cF \rho, \cF \sigma),
$
where $\cF$ ranges over POVMs on $\cH$.
In fact, the infimum above can be attained by a complete orthogonal
measurement on $\cH$.
\end{fact}

The most general operation on a density matrix allowed by quantum
mechanics is what is called a {\em completely positive trace preserving
superoperator}, or superoperator for short. Let $\cH, \cK$ be Hilbert
spaces. A superoperator $\cT$ from $\cH$ to $\cK$ maps quantum states
$\rho$ in $\cH$ to quantum states $\cT \rho$ in $\cK$, and is described 
by a finite
collection of linear maps $\{A_1, \ldots, A_l\}$ from $\cH$ to $\cK$
called {\em Kraus operators} such that, 
$\cT \rho = \sum_{i=1}^l A_i \rho A_i^\dagger$.
Unitary transformations, taking partial traces and POVMs are special
cases of superoperators.

We will use the notation $A \geq B$ for Hermitian operators
$A, B$ in the same Hilbert space $\cH$ as
a shorthand for the statement `$A - B$ is positive semidefinite'.
Thus, $A \geq 0$ denotes that $A$ is positive semidefinite.

Let $X$ be a classical random variable. Let $P$ denote the probability
distribution induced by $X$ on its range $\Omega$.
The {\em Shannon entropy} of 
$X$ is defined as 
$H(X) \defeq H(P) \defeq -\sum_{i \in \Omega} P(i) \log P(i)$.
For any $0 \leq p \leq 1$,
the {\em binary entropy} of $p$ is defined as
$H(p) \defeq H((p, 1-p)) = -p \log p - (1-p) \log (1-p)$.
If $A$ is a quantum system
with density matrix $\rho$, then its {\em von Neumann entropy}
$S(A) \defeq S(\rho) \defeq -\Tr \rho \log \rho$. It is obvious that
the von Neumann entropy of a probability distribution equals its
Shannon entropy. 
If $A, B$ are two
disjoint quantum systems, the {\em mutual information} of $A$ and $B$
is defined as $I(A : B) \defeq S(A) + S(B) - S(AB)$; mutual information
of two random variables is defined analogously. 
By a {\em quantum encoding} $M$ of a classical random variable 
$X$ on $m$ qubits, we mean that there is a bipartite quantum system with 
joint density
matrix $\sum_x \Pr[X = x] \cdot \ketbra{x} \otimes \rho_x$, where the first
system is the random variable, the second system is the quantum
encoding and an $x$ in the range of $X$ is encoded by a quantum
state $\rho_x$ on $m$ qubits. The reduced state of the first system
is nothing but the probability distribution 
$\sum_x \Pr[X = x] \cdot \ketbra{x}$
on the range of $X$. The reduced state of the second system is the
{\em average code word} $\rho \defeq \sum_x \Pr[X = x] \cdot \rho_x$.
The mutual information of this encoding is given by
\[
I(X : M) = S(X) + S(M) - S(X M) = 
S(\rho) - \sum_x \Pr[X = x] \cdot S(\rho_x).
\]

We now define
the {\em relative entropy} of a pair of quantum states.
\begin{definition}[Relative entropy]
\label{def:relentropy}
If $\rho, \sigma$ are quantum states in the same 
Hilbert space, their {\em relative entropy}
is defined as
$S(\rho \| \sigma) \defeq \Tr (\rho (\log \rho - \log \sigma))$.
\end{definition}
For probability distributions $P, Q$ on the same sample space $\Omega$,
the above definition reduces to
$S(P \| Q) = \sum_{i \in \Omega} P(i) \log \frac{P(i)}{Q(i)}$.
The following fact lists some useful properties of relative entropy.
Proofs can be found in \cite[Chapter 11]{nielsen:quant}.
The monotonicity property
below is also called {\em Lindblad-Uhlmann monotonicity}.
\begin{fact}
\label{fact:relentprop}
Let $\rho, \sigma$ be density matrices in the same 
Hilbert space $\cH$. Then,
\begin{enumerate}
\item $S(\rho \| \sigma) \geq 0$, with equality iff $\rho = \sigma$;
\item $S(\rho \| \sigma) < +\infty$ iff
      $\supp(\rho) \subseteq \supp(\sigma)$,
      where $\supp(\rho)$ denotes the {\em support} of $\rho$ i.e.
      the span of the eigenvectors corresponding to non-zero
      eigenvalues of $\rho$;
\item $S(\cdot \| \cdot)$ is continuous in its two arguments when
      it is not infinite.
\item ({\em Unitary invariance}) If $U$ is a unitary transformation on
      $\cH$, 
      $S(U \rho U^\dagger \| U \sigma U^\dagger) = S(\rho \| \sigma)$.
\item ({\em Monotonicity}) Let $\cL$ be a Hilbert space and $\cT$ 
      be a completely positive trace preserving
      superoperator from $\cH$ to $\cL$. Then,
      $S(\cT \rho \| \cT \sigma) \leq S(\rho \| \sigma)$.
\end{enumerate}
\end{fact}

The following fact relates mutual information to relative entropy, and
is easy to prove.
\begin{fact}
\label{fact:inforelentropy}
Let $X$ be a classical random variable and
$M$ be a quantum encoding of $X$ i.e. each $x$ in the range of $X$
is encoded by a quantum state $\rho_x$. Let
$\rho \defeq \sum_x \Pr[X = x] \cdot \rho_x$ be the average code word.
Then, $I(X:M) = \sum_x \Pr[X = x] \cdot S(\rho_x \| \rho)$.
\end{fact}

The next fact is an extension of the random access code arguments
of \cite{ANTV:index}, and was proved by Gavinsky, Kempe, Regev
and de Wolf~\cite[Lemma 1]{deWolfKempe}. 
\begin{fact}
\label{fact:randomaccess}
Let $X = X_1 \cdots X_n$ be a classical random variable of $n$ uniformly
distributed bits. Let $M$ be a quantum encoding of $X$ on $m$ qubits. For
each $i \in [n]$, suppose there is a POVM $\cF_i$ on $M$ with 
three outcomes
$0, 1, ?$. Let $Y_i$ denote the random variable obtained by applying
$F_i$ to $M$. Suppose there are real numbers 
$0 \leq \lambda_i, \epsilon_i \leq 1$ such that
$\Pr[Y_i \neq {} ?] \geq \lambda_i$ and 
$\Pr[Y_i = X_i \mid Y_i \neq {} ?] \geq 1/2 + \epsilon_i$, where the
probability arises from the randomness in $X$ as well as the randomness
of the outcome of $\cF_i$. Then,
\[
\sum_{i=1}^n \lambda_i \epsilon_i^2 \leq
\sum_{i=1}^n \lambda_i (1 - H(1/2 + \epsilon_i)) \leq 
I(X : M) \leq m.
\]
\end{fact}

\section{Privacy trade-offs for set membership}
\label{sec:tradeoffs}
In this section, we prove a trade-off between privacy loss of Alice 
and privacy loss of Bob for the set membership 
problem $\set_n$ assuming the substate theorem.
We then embed index function into other functions
using the concept of VC-dimension and show privacy trade-offs for some
other problems. But first, we 
formally define our model of privacy loss in quantum
communication protocols.

\subsection{Quantum communication protocols}
\label{subsec:privacy}
We consider two party quantum communication
protocols as defined by Yao~\cite{yao:quantcc}. Let $\cX, \cY, \cZ$
be sets and
$f: \cX \times \cY \rightarrow \cZ$ be a function.
There are two players Alice and Bob, who hold qubits. Alice gets
an input $x \in \cX$ and Bob an input $y \in \cY$.
When the communication protocol $\cP$ starts, Alice and Bob each hold
some `work qubits' initialised in the state $\ket{0}$.
Alice and Bob may also share an input independent prior entanglement.
Thus, the initial superposition is simply
$\ket{0}_A \ket{\psi} \ket{y}_B \ket{0}_B$, where
$\ket{\psi}$ is a pure state providing the input independent
prior entanglement.
Here the subscripts denote
the ownership of the qubits by Alice and Bob. Some of the qubits
of $\ket{\psi}$ belong to Alice, the rest belong to Bob.
The players take turns
to communicate to compute $f(x,y)$. Suppose it is Alice's turn.
Alice can make an
arbitrary unitary transformation on her qubits depending on $x$ only
and then send some qubits to Bob. 
Sending qubits does not change the overall
superposition, but rather the ownership of the qubits, allowing
Bob to apply his next unitary transformation, which depends on $y$ only,
on his original qubits
plus the newly received qubits. At the end of the protocol, the
last recipient of qubits performs a measurement in the computational
basis of some qubits in her possession to output an answer
$\cP(x, y)$. For each $(x,y) \in \cX \times \cY$ the unitary 
transformations that are applied, as well as the qubits that 
are to be sent in each round, the number of rounds, the choice of the
starting player,
and the designation of which qubits are to be treated as
`answer qubits' are specified in advance by the protocol $\cP$.
We say that $\cP$ computes $f$ with $\epsilon$-error
in the worst case,
if $\max_{x,y} \Pr[\cP(x, y) \neq f(x,y)] \leq \epsilon$.
We say that $\cP$ computes $f$ with $\epsilon$-error
with respect to a probability distribution
$\mu$ on $\cX \times \cY$, if
$\E_\mu [\Pr [\cP(x, y) \neq f(x,y)]] \leq \epsilon$.
The communication complexity of $\cP$ is defined to be the total number
of qubits exchanged. Note that seemingly more general models of
communication protocols can be thought of, where superoperators may
be applied by the parties instead of unitary transformations and
arbitrary POVM to output the answer of the protocol instead of
measuring in the computational basis, but such
models can be converted to the unitary model above without changing
the error probabilities, communication complexity, and as we will see
later, privacy loss to a cheating party.

Given a probability distribution $\mu$ on $\cX \times \cY$
we define
$
\ket{\mu} \defeq 
\sum_{(x,y) \in \cX \times \cY} \sqrt{\mu(x,y)} \, \ket{x} \ket{y}.
$
Running protocol $\cP$ with superposition $\ket{\mu}$ fed to Alice's and
Bob's inputs means that we first create the state
$
\sum_{(x,y) \in \cX \times \cY} \sqrt{\mu(x,y)}
\ket{x} \ket{0}_A \ket{\psi} \ket{0}_B \ket{y},
$
then feed the middle three registers to $\cP$ and let $\cP$ run
its course till just before applying the final measurement to determine 
the answer of the protocol.
We define the success probability of $\cP$ when 
$\ket{\mu}$ is fed to Alice's and Bob's inputs to be the
probability that measuring the inputs and the answer qubits
in the computational basis at the end of $\cP$ produces consistent results.
Similarly, running protocol $\cP$
with mixture $\mu$ fed to Alice's and Bob's inputs is defined in the
straightforward fashion.
It is easy to see that the success
probability of $\cP$ on superposition $\ket{\mu}$ is the same as
the success probability on mixture $\mu$, that is, the success probability
on superposition $\ket{\mu}$ is equal to
$E_\mu [\Pr [\cP(x, y) = f(x, y)]]$. 

Now let $\mu_{\cX}, \mu_{\cY}$ be probability distributions
on $\cX, \cY$, and let $\mu \defeq \mu_{\cX} \times
\mu_{\cY}$ denote the product distribution on
$\cX \times \cY$.  Let $\cP$ be the prescribed {\em honest} protocol
for $f$. 
Now let us suppose that Bob turns `malicious' and deviates from the
prescribed protocol $\cP$ in order to learn
as much as he can about Alice's input. Note that Alice remains honest
in this scenario i.e. she continues to follow $\cP$.
Thus, Alice and Bob are now actually running a `cheating'
protocol $\widetilde{\cP}$. 
Let registers $A, X, B, Y$ denote Alice's
work qubits, Alice's input qubits, Bob's work qubits and Bob's input
qubits respectively at the end of $\widetilde{\cP}$.
The {\em privacy leakage} from Alice to
Bob in $\widetilde{\cP}$ is captured by the mutual information 
$\widetilde{I}(X : B Y)$
between Alice's input register and Bob's qubits in
$\widetilde{\cP}$. We want to study how large $\sup
\widetilde{I}(X : B Y)$ can be for a given function
$f$, product distribution $\mu$, and protocol $\cP$, where the
supremum is taken over all `cheating' protocols $\widetilde{\cP}$ wherein
Bob can be arbitrarily malicious but Alice continues to follow $\cP$
honestly. We shall call this quantity the {\em privacy loss} of $\cP$ from
Alice to Bob. Privacy leakage and privacy loss from Bob to Alice can
be defined similarly.

One of the ways that Bob can cheat (even without Alice realising it!) 
is by running $\cP$ with the superposition 
$
\ket{\mu_{\cY}} \defeq 
\sum_{y \in \cY} \sqrt{\mu_{\cY}(y)} \, \ket{y}$ 
fed to register $Y$. 
This method of cheating gives Bob at least as much 
information about Alice's input as in the `honest' run of
$\cP$ when the mixture $\mu_{\cY}$ is fed to $Y$.
Sometimes it can give much
more. Consider the set membership problem, where 
Alice has a bit string $x$ which denotes the characteristic vector
of a subset of $[n]$ and Bob has an $i \in [n]$. Consider
a {\em clean} protocol $\cP$ for the index function problem. 
Recall that a protocol $\cP$ is said to be clean if the work
qubits of both the players except the answer qubits are in the state 
$\ket{0}$ at the end of $\cP$. 
We shall show a privacy trade-off result for $\cP$ under the uniform
distribution on the inputs of the two players.
For simplicity, assume that $\cP$ is errorless (an error of
$1/4$ will only change the privacy losses by a multiplicative
constant).
Alice can cheat by feeding a uniform superposition over bit strings
into her input register $X$, and then running $\cP$.
Bob is honest, and has a random $i \in [n]$.
At the end of this `cheating' run of $\cP$, Alice applies a Hadamard 
transformation on each of the registers
$X_j, 1 \leq j \leq n$. Suppose she were to measure them now in
the computational basis. For all $j \neq i$,
she would measure $\ket{0}$ with probability $1$.
For $j = i$, she would measure $1$ with probability $1/2$. Thus,
Alice has extracted about $\log n / 2$ bits of information about
Bob's index $i$. An `honest' run of $\cP$ would have yielded
Alice only $1$ bit of information about $i$. 
Klauck~\cite{klauck:privacy}, based on 
Cleve et al.~\cite{cleve:ip},
has made a similar observation about $\Omega(n)$ privacy loss for
clean protocols computing
the inner product mod $2$ function.
The significance of our lower bounds on privacy loss is
that they make {\em no assumptions} about the protocol $\cP$.

We now define a 
{\em superpositional privacy loss} inspired by the above example. 
We consider a `cheating' run of $\cP$ when mixture 
$\mu_{\cX}$ is fed
to register $X$ and superposition
$\ket{\mu_{\cY}}$
to register $Y$. Let $I'(X : B Y)$ denote
the mutual information of Alice's input register $X$ with
Bob's registers $B Y$ at the
end of this `cheating' run of $\cP$.
\begin{definition}[Superpositional privacy loss]
\label{def:privacy}
The superpositional privacy loss of $\cP$ for function $f$ on the 
product distribution $\mu$ from Alice
to Bob is defined as $L^{\cP}(f, \mu, A, B) \defeq I'(X : B Y)$.
The superpositional privacy loss from Bob to Alice, 
$L^{\cP}(f, \mu, B, A)$,
is defined similarly. 
The superpositional privacy loss of $\cP$ for $f$, $L^{\cP}(f)$, is 
the maximum over 
all product distributions $\mu$, of  
$\max\{L^{\cP}(f, \mu, A, B), L^{\cP}(f, \mu, B, A)\}$.
\end{definition}
\paragraph{Remarks:} \ \\
1.\ \ Our notion of superpositional privacy loss can be viewed 
as a quantum analogue
of the ``combinatorial-informational'' bounded error measure
of privacy loss, $I_{{\rm c-i}}^\ast$, in Bar-Yehuda 
et. al~\cite{baryehuda:privacy}. \\
2.\ \ In \cite{klauck:privacy}, Klauck defines 
a similar notion of privacy loss. In his definition,
a mixture according to
distribution $\mu$ (not necessarily a product distribution)
is fed to both Alice's and Bob's input registers. He does
not consider the case of superpositions being fed to input registers.
For product distributions, our notion of privacy is  
more stringent than
Klauck's, and in fact, the $L^{\cP}(f, \mu, A, B)$ defined above is
an upper bound (to within an additive factor of $\log |\cZ|$) on Klauck's 
privacy loss function. \\
3.\ \ We restrict ourselves to product distributions because we
allow Bob to cheat by putting a superposition in his input register
$Y$.  He should be able to do this
without any {\em a priori} knowledge of $x$, which implies that
the distribution $\mu$ should be a product distribution.
4.\ \ The (general) privacy loss defined above is trivially an upper bound
on the superpositional privacy loss.

\subsection{The privacy trade-off result}
\begin{theorem}
\label{thm:index}
Consider a quantum protocol $\cP$ for $\set_n$ where Alice is given
a subset of $[n]$ and Bob an element of $n$. 
Let $\mu$ denote the uniform 
probability distribution on Alice's and Bob's inputs. 
Suppose $\cP$ has error at most $1/2-\epsilon$ with respect to $\mu$. 
Suppose $L^{\cP}(\set_n, \mu, B, A) \leq k$. Then,
\[
L^{\cP}(\set_n, \mu, A, B) 
\geq \frac{n}{2^{\epsilon^{-3}(14 k + 24)}} - 2.
\]
\end{theorem}
\begin{proof}
Let registers $A, X, B, Y$ denote Alice's work qubits,
Alice's input qubits, Bob's work qubits and Bob's input qubits
respectively, at the end of protocol $\cP$.
We can assume without loss of generality that the last round 
of communication in $\cP$ is from Alice to
Bob, since otherwise, we can add an extra round
of communication at the end wherein Alice sends
the answer qubit to Bob. This process increases 
$L^{\cP}(\set_n, \mu, A, B)$ by at most two 
and does not increase $L^{\cP}(\set_n, \mu, B, A)$
(see e.g. the information theoretic arguments in~\cite{cleve:ip}).
Thus at the end of $\cP$, Bob measures the answer qubit,
which is a qubit in the register $B$,
in the computational basis to determine 
$f(x, y)$. In the proof, subscripts of pure
and mixed states will denote the registers which are in those
states.

Let $\ket{\psi_i}_{X A Y B}$ be the state vector of Alice's and 
Bob's qubits
and $(\rho_i)_{X A}$ the density matrix of Alice's qubits at the end
of the protocol $\cP$, when Alice is fed a uniform superposition
over bit strings
in her input register $X$ and Bob is fed $\ket{i}$ in 
his input register $Y$. Let $1/2 + \epsilon_i$ be the success 
probability of $\cP$ in this case. Without loss of generality,
$\epsilon_i \geq 0$.
Consider a run, Run~1, of $\cP$ when a uniform mixture
of indices is fed to register $Y$, and a uniform superposition over
bit strings is fed to register $X$. Let $1/2 + \epsilon$ be
the success probability of $\cP$ for Run~1, which is also 
the success probability of $\cP$ with respect to $\mu$.  Then
$1/4 \leq \epsilon = (1/n) \sum_{i=1}^n \epsilon_i$. 
Let $I_1(Y : A X)$ denote
the mutual information of register $Y$ with registers $A X$ at the
end of Run~1 of $\cP$. We know that 
$I_1(Y : A X) = L^{\cP}(\set_n, \mu, B, A) \leq k$.
Let $\rho_{XA} \defeq (1/n) \sum_{i=1}^n (\rho_i)_{XA}$ and
$k_i \defeq S((\rho_i)_{XA} \| \rho_{XA} )$. 
Note that $0 \leq k_i < \infty$ by Fact~\ref{fact:relentprop}.
By Fact~\ref{fact:inforelentropy}, 
\begin{displaymath}
k \geq I_1(Y : A X) = \frac{1}{n} \sum_{i=1}^n 
                      S((\rho_i)_{XA} \| \rho_{XA}) 
                    = \frac{1}{n} \sum_{i=1}^n k_i.
\end{displaymath}
Let $k_i' \defeq k_i + 4 \sqrt{k_i + 2} + 2 \log (k_i + 2) + 5$ and
$r_i \defeq (2/\epsilon_i)^2$. 

Let us now consider a run, Run~2, of $\cP$
with uniform superpositions fed to registers $X, Y$. 
Let $\ket{\phi}_{X A Y B}$ be the state vector of Alice's 
and Bob's qubits at the end of Run~2 of $\cP$. Then,
$\parttr{YB} \ketbra{\phi} = \rho_{XA}$, and 
the success probability of $\cP$ for Run~2 is 
$1/2 + \epsilon$.
Let $Q$ be an additional qubit.
By the substate theorem (Theorem~\ref{thm:substate}), there exist states 
$\ket{\psi'_i}_{XAYBQ}$, $\ket{\theta'_i}_{XAYBQ}$ such that
$\trnorm{\ketbra{\psi_i} - \ketbra{\psi_i'}} 
\leq 2/\sqrt{r_i} = \epsilon_i$ and 
$\parttr{YBQ} \ketbra{\phi_i} = \rho_{XA}$ where
\begin{displaymath}
\ket{\phi_i}_{X A Y B Q} \defeq
     \sqrt{\frac{r_i - 1}{r_i 2^{r_i k_i'}}} \,
           \ket{\psi_i'}_{X A Y B} \ket{1}_{Q} +
     \sqrt{1 - \frac{r_i - 1}{r_i 2^{r_i k_i'}}} \,
           \ket{\theta_i'}_{X A Y B} \ket{0}_{Q},
\end{displaymath}
In fact, there exists
a unitary transformation $U_i$ on registers $Y B Q$,
transforming the state $\ket{\phi}_{X A Y B} \ket{0}_{Q}$ 
to the state $\ket{\phi_i}_{X A Y B Q}$.

For each $i \in [n]$, let $X_i'$ denote the classical random variable
got by measuring the $i$th bit of register $X$ in state
$\ket{\phi}_{X A Y B}$.
We now prove the following claim. 
\begin{claim}
For each $i \in [n]$,
there is a POVM ${\cal M}_i$ with three outcomes $0$, $1$, $?$ acting
on $Y B$ such that if $Z_i'$ is the result of ${\cal M}_i$ on 
$\ket{\phi}_{X A Y B}$, then 
$
\Pr[Z_i' \neq \; ?] 
\geq 2^{-4 \epsilon_i^{-2} (k'_i+1)}, 
$
and
$\Pr[Z_i' = X_i' \mid Z_i' \neq \; ?] \geq 1/2 + \epsilon_i/2$. 
\end{claim}
\begin{proof}
The POVM ${\cal M}_i$ proceeds by first bringing in the ancilla qubit
$Q$ initialised to $\ket{0}_Q$, then applying $U_i$ to the registers 
$Y B Q$ and finally
measuring $Q$ in the computational basis. 
If it observes $\ket{1}_Q$, ${\cal M}_i$ measures 
the answer qubit in $B$ in the computational basis and declares
the result as $Z_i'$. If it observes $\ket{0}_Q$, ${\cal M}_i$ outputs $?$.

When applied to $\ket{\phi}_{X A Y B}$, ${\cal M}_i$ first generates
$\ket{\phi_i}_{X A Y B Q}$ and then measures $Q$ in the computational
basis.
In the case when ${\cal M}_i$ measures $\ket{1}$ for qubit $Q$, which 
happens with probability 
\[
\Pr[Z_i' \neq \; ?] 
  =  \frac{r_i - 1}{r_i 2^{r_i k_i'}}
\geq 2^{-4 \epsilon_i^{-2} (k'_i+1)}, 
\]
the state vector of $X A Y B$ collapses to
$\ket{\psi_i'}$. In this case by Fact~\ref{fact:totvartrace},
\begin{displaymath}
\Pr[Z_i' = X_i' | Z_i' \neq \; ?] 
   \geq \frac{1}{2} + \epsilon_i - \frac{1}{2} 
        \trnorm{\ketbra{\psi_i} - \ketbra{\psi_i'}}
   \geq \frac{1}{2} + \frac{\epsilon_i}{2}.
\end{displaymath}
\end{proof}

Consider now a run, Run~3, of $\cP$ when a uniform mixture over
bit strings is fed to register $X$ and a uniform superposition
over $[n]$ is fed to register $Y$. Let $\rho_{X A Y B}$ denote
the density matrix of the registers $X A Y B$ at the end of
Run~3 of $\cP$. In fact, measuring in the 
computational basis the register $X$ in the
state $\ket{\phi}_{X A Y B}$ gives us $\rho_{X A Y B}$; also,
$\parttr{YB} \rho_{X A Y B} = \rho_{XA}$.
Let $I_3(X : Y B)$ denote the mutual information between register
$X$ and registers $Y B$ in the state $\rho_{X A Y B}$.
For each $i \in [n]$, let $X_i$ denote the classical 
random variable corresponding to the $i$th bit of register $X$ in state
$\rho_{X A Y B}$. Then,
$X \defeq X_1 \ldots X_n$ is a uniformly distributed bit string of
length $n$. 
Let $Z_i$ denote the result of POVM ${\cal M}_i$ of the above claim applied
to $\rho_{X A Y B}$. Then since ${\cal M}_i$ acts only on the registers
$Y B$, we get
$
\Pr[Z_i \neq \; ?] 
  =  \Pr[Z_i' \neq \; ?] 
\geq 2^{-4 \epsilon_i^{-2} (k_i' + 1)},
$
and
$\Pr[Z_i = X_i \mid Z_i \neq \; ?] 
  =  \Pr[Z_i' = X_i' \mid Z_i \neq \; ?] 
\geq 1/2 + \epsilon_i/2$.
Define 
$
{\sf Good} \defeq 
\{i \in [n]: k_i \leq 2 k/\epsilon, \epsilon_i \geq \epsilon/2\}.
$
By Markov's inequality, $|{\sf Good}| > n \epsilon/2$.
By Fact~\ref{fact:randomaccess},
\begin{eqnarray*}
I(X : Y B) 
& \geq & \sum_{i=1}^n 
         \frac{\epsilon_i^2\cdot 2^{-4 \epsilon_i^{-2}(k_i'+1)}}{4}
\;\geq\; \sum_{i \in {\sf Good}} 
         \frac{\epsilon_i^2\cdot 2^{-4 \epsilon_i^{-2}(k_i'+1)}}{4} \\
& \geq & \frac{n\epsilon^3\cdot 
               2^{\epsilon^{-3}(2k+4\sqrt{2k+2}+2\log(2k+2)+6)}
              }
              {32} 
\;\geq\; \frac{n}{2^{\epsilon^{-3}(2k+4\sqrt{2k+2}+2\log(2k+2)+12)}} \\
& \geq & \frac{n}{2^{\epsilon^{-3}(14 k + 24)}}.
\end{eqnarray*}
By the arguments in the first paragraph of this proof, we have
$L^{\cP}(\set_n, \mu, A, B) \geq I(X : Y B) - 2$. This 
completes the proof of the theorem.
\end{proof}\\
{\bf Remark:} 
This theorem is the formal version of Result~\ref{res:privacy} stated
in the introduction. 

As we have mentioned earlier, this theorem has been generalised 
in~\cite{jain:entangred} in a suitable manner to relate the  privacy
loss for any function in terms of its one-way communication
complexity. We do not get into the details of this statement here.
Instead, we give a weaker corollary of the present theorem that relates
the privacy loss of a function to the {\em Vapnik-Chervonenkis dimension
(VC-dimension)} of its communication matrix.
\begin{definition}[VC-dimension]
For a boolean valued function 
$f: \cX \times \cY \rightarrow \{0, 1\}$, a set
$T \subseteq \cY$ is {\em shattered}, if for all $S \subseteq T$
there is an $x \in \cX$ such that 
$\forall y \in T: f(x, y) = 1 \Leftrightarrow
y \in S$. The {\em VC-dimension} of $f$ for 
$\cX$, $\vc_{\cX}(f)$, is the largest size of such a 
shattered set $T \subseteq \cY$. We define $\vc_{\cY}(f)$ analogously.
\end{definition}
Informally, $\vc_{\cX}(f)$ captures the size of the largest instance of
the set membership problem $\set_n$ that can be `embedded' into $f$.
Using this connection, one can trivially prove a privacy trade-off
result for $f$ in terms of $\vc_{\cX}(f)$, $\vc_{\cY}(f)$ by
invoking Theorem~\ref{thm:index}. This generalises
Klauck's lower bound~\cite{klauck:oneround} for 
the communication complexity of bounded error one-way quantum protocols
for $f$ in terms of its VC-dimension. 
\begin{corollary}
\label{cor:vcdim}
Let $f: \cX \times \cY \rightarrow \{0, 1\}$ be a
boolean valued function. Let $\vc_{\cX}(f) = n$. 
Then there is a product distribution $\mu$ on 
$\cX \times \cY$ such that,
if $\cP$ is a quantum protocol for $f$ with
average error at most $1/2 - \epsilon$ with respect to $\mu$, 
\[
L^{\cP}(f, \mu, B, A) \leq k 
\Leftrightarrow
L^{\cP}(f, \mu, A, B) \geq 
\frac{n}{2^{\epsilon^{-3}(14 k + 24)}} - 2.
\]
An analogous statement holds for $\vc_{\cY}(f)$.
\end{corollary}
\begin{proof} 
Since $\vc_{\cX}(f) = n$, there is a set
$T \subseteq \cY$, $|T| = n$ which is shattered. Without
loss of generality, $T = [n]$. For any subset $S \subseteq T$,
there is an $x \in \cX$ such that 
$\forall y \in T: f(x, y) = 1 \Leftrightarrow
y \in S$. We now give a reduction from $\set_n$
to $f$ as follows: In $\set_n$, Alice is given a subset $S \subseteq [n]$
and Bob is given a $y \in [n]$. Alice and Bob run the protocol
$\cP$ for $f$ on inputs $x$ and $y$ respectively, to solve
$\set_n$. The corollary now follows from Theorem~\ref{thm:index}.
\end{proof}

The following consequence of Corollary~\ref{cor:vcdim} is immediate.
\begin{corollary}
\label{cor:privacy}
Quantum protocols for
set membership $\set_n$, set disjointness for subsets of $[n]$ and 
inner product modulo $2$ in $\{0,1\}^n$ each
suffer from $\Omega(\log n)$ privacy loss.
\end{corollary}
\begin{proof}
Follows trivially from Corollary~\ref{cor:vcdim} since all the three
functions have VC-dimension $n$.
\end{proof}

\section{The substate theorem}
\label{sec:substate}
In this section, we prove the quantum substate theorem. But first, we
state a fact from game theory that will be used in its proof. 

\subsection{A minimax theorem}
We will require the following minimax theorem from game theory, which
is a consequence of the Kakutani fixed point theorem in real
analysis.
\begin{fact}
\label{fact:minimax}
Let $A_1,A_2$ be non-empty, convex and compact subsets of $\reals^n$
for some $n$. Let $u: A_1 \times A_2 \rightarrow \reals$ be a
continuous function, such that 
\begin{itemize}
\item $\forall a_2 \in A_2$, the set 
$\{a_1 \in A_1:\forall a_1' \in A_1 \, 
               u(a_1,a_2) \geq u(a_1',a_2)\}$ is convex; and
\item  $\forall a_1 \in A_1$, the set 
$\{a_2 \in A_2: \forall a_2' \in A_2 \, 
                u(a_1,a_2) \leq u(a_1,a_2')\}$ is convex.
\end{itemize}
Then, there is an $(a_1^\ast, a_2^\ast) \in A_1 \times A_2$ such that 
\begin{displaymath}
\max_{a_1\in A_1}\, \min_{a_2\in A_2} u(a_1,a_2) 
= u(a_1^\ast, a_2^\ast)
= \min_{a_2 \in A_2}\, \max_{a_1 \in A_1} u(a_1,a_2).
\end{displaymath}
\end{fact}
\paragraph{Remark:}
The above statement follows by combining Proposition~20.3 
(which shows the
existence of Nash equilibrium $a^\ast$ in strategic games) and
Proposition~22.2  (which connects Nash equilibrium and the min-max
theorem for games defined using a pay-off function such as $u$) of
Osborne and Rubinstein's~\cite[pages 19--22]{osborne:gametheory} 
book on game theory.

\subsection{Proof of the substate theorem}
\label{subsec:substate}
We now state the quantum substate theorem as it is actually used in
our privacy lower bound proofs.
\begin{theorem}[Quantum substate theorem]
\label{thm:substate}
Consider two Hilbert spaces $\cH$ and $\cK$,
$\dim(\cK) \geq \dim(\cH)$. Let $\complexes^2$ denote the
two dimensional complex Hilbert space.
Let $\rho, \sigma$ be density matrices in $\cH$.
Let $r > 1$ be any real number. Let $k \defeq S(\rho \| \sigma)$.
Let $\ket{\psi}$
be a purification of $\rho$ in $\cH \otimes \cK$. Then there
exist pure states
$\ket{\phi}, \ket{\theta} \in \cH \otimes \cK$ and
$\ket{\zeta} \in \cH \otimes \cK \otimes \complexes^2$,
depending on $r$, 
such that $\ket{\zeta}$ is a purification of $\sigma$ and
$\trnorm{\ketbra{\psi} - \ketbra{\phi}} \leq 2/\sqrt{r}$, where
\begin{displaymath}
\ket{\zeta} \defeq
     \sqrt{\frac{r - 1}{r 2^{r k'}}} \, \ket{\phi}\ket{1} +
     \sqrt{1 - \frac{r - 1}{r 2^{r k'}}} \, \ket{\theta}\ket{0}
~~~ {\rm and} ~~~
k' \defeq k + 4 \sqrt{k+2} + 2 \log (k+2) + 5. 
\end{displaymath}
\end{theorem}
{\bf Remarks:} \\
1.\ \ Note that Result~\ref{res:substate}
      in the introduction follows from above by 
      tracing out $\cK \otimes \complexes^2$. \\
2.\ \ From Result~\ref{res:substate}, one can easily see that 
      $\trnorm{\rho - \sigma} \leq 2 - 2^{-O(k)}$. This implies a
      $2^{-O(k)}$ lower bound on the fidelity of $\rho$ and $\sigma$. 

\bigskip

\noindent
{\bf Overview of the proof of Theorem~\ref{thm:substate}:} As we 
have mentioned earlier, our proof of
the quantum substate theorem goes through first by
defining a new notion of
distinguishability called {\em observational divergence},
$D(\rho \| \sigma)$, between two density matrices $\rho$, $\sigma$ in the
same Hilbert space $\cH$. Informally speaking, this notion is a single
observational version of relative entropy. Truly speaking,
the substate theorem is a
relationship between observational divergence and the substate condition. 
We first prove an observational divergence lifting theorem which shows 
that given two
states $\rho, \sigma$ in $\cH$ and any extension $\sigma'$ of $\sigma$
in $\cH \otimes \cK$, $\dim(\cK) \geq \dim(\cH)$, one can find a 
purification $\ket{\phi}$ of
$\rho$ in $\cH \otimes \cK$ such that 
$D(\ketbra{\phi} \| \sigma) = O(D(\rho \| \sigma))$.
This theorem may be of independent
interest. This helps us reduce the statement we intend to prove only
to the case when $\rho$ is a pure state. This case is then further reduced
to analysing only a two dimensional scenario which is then resolved by a
direct calculation. The final statement of the quantum substate theorem
in terms of relative entropy
is established by showing that observational divergence is never much
bigger than relative entropy for any pair of states.

Let us begin by defining observational divergence. 
\begin{definition}[Observational divergence]
\label{def:div}
Let $\rho, \sigma$ be density matrices in the same 
Hilbert space $\cH$. Their observational divergence
is defined as
\begin{displaymath}
D(\rho \| \sigma) \defeq
    \sup_F \left(\Tr (F \rho) \log \frac{\Tr (F \rho)}{\Tr (F \sigma)}
           \right),
\end{displaymath}
where $F$ above ranges over POVM elements on $\cH$
such that $\Tr (F \sigma) \neq 0$.
\end{definition}

The following properties of observational
divergence follow easily from the definition.
\begin{proposition}
\label{prop:propdiv}
Let $\rho, \sigma$ be density matrices in the same 
Hilbert space $\cH$. Then
\begin{enumerate}
\item $D(\rho \| \sigma) \geq 0$, with equality iff $\rho = \sigma$.
\item $D(\rho \| \sigma) < +\infty$ iff
      $\supp(\rho) \subseteq \supp(\sigma)$.
      If $D(\rho \| \sigma) < +\infty$, then
      there is a POVM element $F$ which achieves equality in
      Definition~\ref{def:div}.
\item $D(\cdot \| \cdot)$ is continuous in its two arguments when
      it is not infinite.
\item {\em (Unitary invariance)} If $U$ is a unitary transformation on
      $\cH$, 
      $D(U \rho U^\dagger \| U \sigma U^\dagger) = D(\rho \| \sigma)$.
\item {\em (Monotonicity)} Suppose $\cK$ is a Hilbert
      space, and $\rho', \sigma'$ are extensions of $\rho, \sigma$ in
      $\cH \otimes \cK$. Then,
      $D(\rho' \| \sigma') \geq D(\rho \| \sigma)$. This implies,
      via unitary invariance and the Kraus representation theorem,
      that if $\cT$ is a completely positive trace preserving
      superoperator from $\cH$ to a Hilbert space $\cL$, then
      $D(\cT \rho \| \cT \sigma) \leq D(\rho \| \sigma)$.
\end{enumerate}
\end{proposition}

Fact~\ref{fact:relentprop} and Proposition~\ref{prop:propdiv} seem 
to suggest
that relative entropy and observational divergence are 
similar quantities. In fact, the relative entropy is an upper bound
on the observational divergence to within an additive constant.
More properties of observational divergence as well as comparisons
with relative entropy are discussed in the appendix.
\begin{proposition}
\label{prop:divrelentropy}
Let $\rho, \sigma$ be density matrices in the same 
Hilbert space $\cH$. Then,
$D(\rho \| \sigma ) < S(\rho \| \sigma) + 1$.
\end{proposition}
\begin{proof}
By Fact~\ref{fact:relentprop} and
Proposition~\ref{prop:propdiv}, 
$D(\rho \| \sigma) = +\infty$
iff $\supp(\rho) \not \subseteq \supp(\sigma)$ iff
$S(\rho \| \sigma) = +\infty$. Thus,
we can henceforth assume without loss of generality that
$D(\rho \| \sigma) < +\infty$. 
By Proposition~\ref{prop:propdiv},
there is a POVM element $F$ such that
$D(\rho \| \sigma) = p \log (p/q)$,
where $p \defeq \Tr (F \rho)$ and $q \defeq \Tr (F \sigma)$.
We now have
\begin{eqnarray*}
S(\rho \| \sigma )
& \geq & p \log \frac{p}{q} +
           (1 - p) \log \frac{(1 - p)}{(1 - q)} 
\;  > \; p \log \frac{p}{q} +
           (1 - p) \log \frac{1}{(1 - q)} - 1 
\;\geq\; p \log \frac{p}{q} - 1 \\
&   =  & D(\rho \| \sigma) - 1.
\end{eqnarray*}
The first inequality follows from the Lindblad-Uhlmann monotonicity
of relative entropy (Fact~\ref{fact:relentprop}), 
and the second
inequality follows because
$(1 - p) \log (1 - p) \geq (- \log e)/e > -1$,
for $0 \leq p \leq 1$. This completes the proof of the lemma.
\end{proof}

We now prove the following lemma, which can be thought of as a
substate theorem when the first density matrix is in fact
a pure state. 
\begin{lemma}
\label{lem:twobytwo}
Let $\ket{\psi}$ be a pure state and $\sigma$ be a density matrix
in the same Hilbert space $\cH$.
Let $k \defeq D\left((\ketbra{\psi}) \| \sigma\right)$.
Then for all $r \geq 1$, there exists a pure state
$\ket{\phi}$, depending on $r$, such that
\begin{displaymath}
\trnorm{\ketbra{\psi} - \ketbra{\phi}} < \frac{2}{\sqrt{r}}
~~~ {\rm and} ~~~
\left(\frac{r-1}{r 2^{rk}}\right) \ketbra{\phi} < \sigma.
\end{displaymath}
\end{lemma}
\begin{proof}
We assume without loss of generality that $0 < k < +\infty$.
Consider $M \defeq \sigma - (\ketbra{\psi}/2^{rk})$. Since
$-(\ketbra{\psi}/2^{rk})$ has exactly one non-zero eigenvalue and this
eigenvalue is negative viz. $-1/2^{rk}$, and $\sigma$ is positive
semidefinite, $M$ is a
hermitian matrix with at most one negative eigenvalue.

If $M \geq 0$ we take $\ket{\phi}$ to be $\ket{\psi}$. The
lemma trivially holds in this case.

Otherwise, let $\ket{w}$ be the eigenvector corresponding to
the unique negative eigenvalue $-\alpha$ of $M$.
Thinking of $\ketbra{w}$ as a POVM element, we get
\begin{displaymath}
0 > -\alpha = \Tr (M \ketbra{w}) =
    \unibraket{w}{\sigma} - \frac{|\braket{\psi}{w}|^2}{2^{rk}}
\Rightarrow
    \unibraket{w}{\sigma} < \frac{|\braket{\psi}{w}|^2}{2^{rk}}.
\end{displaymath}
Hence
\begin{displaymath}
k = D(\ketbra{\psi} \| \sigma) \geq
    |\braket{\psi}{w}|^2 \log \frac{|\braket{\psi}{w}|^2}
                                   {\unibraket{w}{\sigma}} >
    r k |\braket{\psi}{w}|^2
\Rightarrow
    |\braket{\psi}{w}|^2 < \frac{1}{r} \leq 1.
\end{displaymath}
In particular, this shows that $\ket{\psi}, \ket{w}$ are linearly
independent.

Let $n \defeq \dim (\cH)$.
Let $\{\ket{v}, \ket{w}\}$ be an orthonormal basis for
the two dimensional subspace of $\cH$ spanned by
$\{\ket{\psi}, \ket{w}\}$. Extend it to
$\{\ket{v_1}, \ldots, \ket{v_{n-2}}, \ket{v}, \ket{w}\}$,
an orthonormal basis
for the entire space $\cH$.
In this basis we have the following matrix equation,
\begin{equation}
\label{eq:matrix}
\threearray{F}{e}{d}{a}{b}{c} - \threearray{0}{0}{0}{x}{y}{z} =
   \bigtwoarray{P}{l}{-\alpha},
\end{equation}
where the first, second and third matrices are $\sigma$,
$\ketbra{\psi}/2^{rk}$ and $M$ respectively.
$F$ is an $(n - 2) \times (n - 2)$ matrix, $P$ is an
$(n - 1) \times (n - 1)$ matrix, $d$, $e$ are $(n - 2) \times 1$
matrices and $l$ is an $(n - 1) \times 1$ matrix.
$a, c, x, z, \alpha$ are non-negative real numbers and $b, y$ are
complex numbers. The zeroes above denote
all zero matrices of appropriate dimensions. The dagger denotes
conjugate transpose.

\begin{claim}
\label{claim:twobytwoineq}
We have the following properties.
\begin{enumerate}
\item $b, y \in \complexes$, $a, c, x, z, \alpha \in \reals$.
\item $b = y \neq 0$, $1/(r2^{rk}) > z = c + \alpha > c > 0$,
      $\alpha > 0$, $a > 0$, $0 < x < 1/2^{rk}$,  $x + z = 1/2^{rk}$,
      $l = 0$ and $d = 0$.
\item $0 < \frac{x c}{|b|^2} < \frac{x z}{|y|^2} = 1$.
\end{enumerate}
\end{claim}
\begin{proof}
The first part of the claim has already been mentioned above.
Since $\ket{w}$ is an eigenvector of $M$ corresponding to eigenvalue
$-\alpha$, $l = 0$.
By inspection, we have $b = y, z = c + \alpha, d = 0$.
We have $x > 0$ since $\ket{\psi}, \ket{w}$ are linearly
independent, and $z > c \geq 0$ since $\alpha > 0$.
Now, $x + z = \Tr (\ketbra{\psi}/2^{rk}) = 1/2^{rk}$ and so
$x < 1/2^{rk}$. Also, $z = |\braket{\psi}{w}|^2/2^{rk} < 1/(r2^{rk})$.
Since $\sigma \geq 0$, $F \geq 0$ and
$\twoarray{a}{b}{c} \geq 0$. Hence,
\begin{displaymath}
\det \twoarray{a}{b}{c} = a c - |b|^2 \geq 0.
\end{displaymath}
Since $\ketbra{\psi}/2^{rk}$ has one dimensional support,
\begin{displaymath}
\det \twoarray{x}{y}{z} = x z - |y|^2 = 0.
\end{displaymath}
If $c = 0$ then $y = b = 0$, which implies that $xz = 0$, which is
a contradiction. Hence, $c > 0$ and $b \neq 0$. Similarly,
$a > 0$. This proves the second part of the claim. The third
part now follows easily.
\end{proof}

We can now write $\sigma = \sigma_1 + \sigma_2$, where
\begin{displaymath}
\sigma_1 \defeq \threearray{F}{e}{0}{a - \frac{|b|^2}{c}}{0}{0}
~~~ {\rm and} ~~~
\sigma_2 \defeq \threearray{0}{0}{0}{\frac{|b|^2}{c}}{b}{c}.
\end{displaymath}
Note that $\ket{\xi} = (0, \ldots, 0, 1, -b^\dag/c)$ is
an eigenvector of $\sigma_2$ corresponding to the eigenvalue $0$.
We have $\sigma_2 \geq 0$, and in fact, $\sigma_2$ has one
dimensional support. We now claim that
$\sigma_1 \geq 0$. For otherwise, since $F \geq 0$, there is a
vector $\ket{\theta}$ of the form
$(a_1, \ldots ,a_{n - 2}, 1, 0)$ such that
$\unibraket{\theta}{\sigma_1} < 0$.
Now consider the vector
$\ket{\theta'} \defeq (a_1, \ldots, a_{n - 2}, 1, -b^\dag/c)$.
We have,
\begin{displaymath}
\unibraket{\theta'}{\sigma} =
      \unibraket{\theta'}{\sigma_1} +
      \unibraket{\theta'}{\sigma_2} =
      \unibraket{\theta}{\sigma_1} +
      \unibraket{\xi}{\sigma_2} < 0,
\end{displaymath}
contradicting $\sigma \geq 0$.
This shows that $\sigma_1 \geq 0$, and hence, $\sigma \geq \sigma_2 $.

We are now finally in a position to define the pure state $\ket{\phi}$.
Note that $\ketbra{\phi}$ is nothing but $\sigma_2$ normalised to have
unit trace. That is,
\[
\ketbra{\phi} \defeq \frac{\sigma_2}{\frac{|b|^2}{c} + c}.
\]

Using Claim~\ref{claim:twobytwoineq} we get,
\begin{displaymath}
\Tr \sigma_2 = \frac{|b|^2}{c} + c
             > \frac{|b|^2}{z} + c
             = x + z - \alpha
             > \frac{r-1}{r 2^{rk}}.
\end{displaymath}
Hence,
$
\frac{r - 1}{r 2^{rk}} \ketbra{\phi} < \sigma_2 \leq \sigma.
$
This shows the second assertion of the lemma.

To complete the proof of the lemma,
we still need to show that
$\trnorm{\ketbra{\psi} - \ketbra{\phi}}$ is small.
Up to global phase factors, one can write $\ket{\psi}, \ket{\phi}$
as follows:
\begin{displaymath}
\ket{\psi} = \frac{\frac{b}{\sqrt{z}} \ket{v} + \sqrt{z} \ket{w}}
                  {\sqrt{\frac{|b|^2}{z} + z}},
~~~~~
\ket{\phi} = \frac{\frac{b}{\sqrt{c}} \ket{v} + \sqrt{c} \ket{w}}
                  {\sqrt{\frac{|b|^2}{c} + c}}.
\end{displaymath}
We now lower bound $|\braket{\phi}{\psi}|$ as follows, 
using Claim~\ref{claim:twobytwoineq}.
\begin{eqnarray*}
|\braket{\phi}{\psi}| 
& = & \frac{\frac{|b|^2}{\sqrt{cz}} + \sqrt{cz}}
           {\sqrt{\frac{|b|^2}{c}+c} \cdot \sqrt{\frac{|b|^2}{z} + z}} 
\;=\; \frac{|b|^2 + cz}{\sqrt{(|b|^2 + c^2)(|b|^2 + z^2)}} \\
& > & \frac{|b|^2 + cz}{\sqrt{(|b|^2 + cz)(|b|^2 + z^2)}} 
\;=\; \sqrt{\frac{|b|^2 + cz}{|b|^2 + z^2}} 
\;=\; \sqrt{\frac{x + c}{x + z}} 
\;=\; \sqrt{1 - \frac{\alpha}{x + z}} \\
& > & \sqrt{1 - \frac{1}{r}}.
\end{eqnarray*}
This proves that
$
\trnorm{\ketbra{\psi} - \ketbra{\phi}} 
= 2 \sqrt{1 - |\braket{\phi}{\psi}|^2} 
< 2/\sqrt{r},
$
establishing the first assertion of the lemma and completing its
proof.
\end{proof}

We next prove the following lemma, which can be thought of as
an `observational substate' lemma.
\begin{lemma}
\label{lem:lifting1}
Consider two Hilbert spaces $\cH$ and $\cK$,
$\dim(\cK) \geq \dim(\cH)$.
Let $\rho, \sigma$ be density matrices in $\cH$.
Let $\ket{\psi}$ be a purification of $\rho$ in $\cH \otimes \cK$.
Let $F$ be a POVM element on $\cH \otimes \cK$. Let $\beta > 1$.
Then there exists
a purification $\ket{\phi}$ of $\sigma$ in $\cH \otimes \cK$
such that
$q \geq \frac{p}{2^{k'/p}}$,
where $p \defeq \Tr (F \ketbra{\psi})$,
$q \defeq \Tr (F \ketbra{\phi})$ and
$k' \defeq \beta D(\rho \| \sigma) - 2 \log (1 - \beta^{-1/2}) $.
\end{lemma}
\begin{proof}
We assume without loss of
generality that $0 < D(\rho \| \sigma) < +\infty$ and that $p > 0$.
Let $n \defeq \dim (\cH \otimes \cK)$ and $\{\ket{\alpha_i}\}_{i=1}^n$ be
the orthonormal eigenvectors of $F$ with corresponding eigenvalues
$\{\lambda_i\}_{i=1}^n$. Note that $0 \leq \lambda_i \leq 1$ and
$\ket{\alpha_i} \in \cH \otimes \cK$.
We have,
\begin{displaymath}
p = \sum_{i=1}^n \lambda_i |\braket{\alpha_i}{\psi}|^2
~~~ {\rm and} ~~~
q = \sum_{i=1}^n \lambda_i |\braket{\alpha_i}{\phi}|^2.
\end{displaymath}
Define,
\begin{displaymath}
\ket{\theta'} \defeq
\frac{\sum_{i=1}^n \lambda_i \braket{\alpha_i}{\psi} \ket{\alpha_i}}
     {\sqrt{p}}
~~~ {\rm and} ~~~
\ket{\theta} \defeq \frac{\ket{\theta'}}{\|\ket{\theta'}\|}.
\end{displaymath}
Note that $p = |\braket{\psi}{\theta}|^2 \|\ket{\theta'}\|^2$ and
$0 < \|\ket{\theta'}\|^2 \leq 1$. Using the
Cauchy-Schwarz inequality, we see that
\begin{displaymath}
|\braket{\phi}{\theta}|^2 \|\ket{\theta'}\|^2 =
|\braket{\phi}{\theta'}|^2 =
\frac{\left| \sum_{i=1}^n \lambda_i \braket{\alpha_i}{\psi}
                          \braket{\phi}{\alpha_i}\right|^2}
     {\sum_{i=1}^n \lambda_i |\braket{\alpha_i}{\psi}|^2} \leq
\sum_{i=1}^n \lambda_i |\braket{\alpha_i}{\phi}|^2 = q.
\end{displaymath}
Thus,
\begin{displaymath}
\frac{p}{2^{k'/p}} =
\frac{|\braket{\psi}{\theta}|^2 \|\ket{\theta'}\|^2}
     {2^{k'/(|\braket{\psi}{\theta}|^2 \|\ket{\theta'}\|^2)}} \leq
\frac{|\braket{\psi}{\theta}|^2 \|\ket{\theta'}\|^2}
     {2^{k'/|\braket{\psi}{\theta}|^2}}.
\end{displaymath}
Hence, it will suffice to show that there exists a purification
$\ket{\phi}$ of $\sigma$ in $\cH \otimes \cK$ such that
\begin{displaymath}
|\braket{\phi}{\theta}|^2 \geq \frac{|\braket{\psi}{\theta}|^2}
                                    {2^{k'/|\braket{\psi}{\theta}|^2}}.
\end{displaymath}

Define the density matrix $\tau$ in $\cH$ as
$\tau \defeq \parttr{\cK} \ketbra{\theta}$.
By Facts~\ref{fact:jozsa} and \ref{fact:fuchscaves}, there is a
purification $\ket{\phi}$ of $\sigma$ in $\cH \otimes \cK$ and a
POVM $\{F_1, \ldots, F_l\}$ in $\cH$ such that,
\begin{displaymath}
|\braket{\phi}{\theta}| =
B(\tau, \sigma) = \sum_{i=1}^l \sqrt{c_i b_i},
\end{displaymath}
where $c_i \defeq \Tr (F_i \tau)$ and $b_i \defeq \Tr (F_i \sigma)$.
Let $a_i \defeq \Tr (F_i \rho)$. We know from
Facts~\ref{fact:jozsa} and \ref{fact:fuchscaves} that
\begin{displaymath}
0 < \sqrt{p} \leq |\braket{\psi}{\theta}| \leq
B(\tau, \rho) \leq \sum_{i=1}^l \sqrt{c_i a_i}.
\end{displaymath}
Note that the $a_i$'s are non-negative real numbers summing up to $1$,
and so are the $b_i$'s and the $c_i$'s.

For $\beta > 1$, define the set 
$
S_\beta \defeq 
\left\{i \in [l]: a_i > b_i \cdot 2^{\beta k/B(\tau, \rho)^2}\right\}$,
where $k \defeq D(\rho \| \sigma)$.
Note that $\forall i \in S, b_i \neq 0$ as $\supp(\rho) \subseteq
\supp(\sigma)$, $k$ being finite.
Define the POVM element $G$ on $\cH$ as
$G \defeq \sum_{i \in S_\beta} F_i$.
Let $a \defeq \Tr (G \rho)$ and $b \defeq \Tr (G \sigma)$. Then
$a = \sum_{i \in S_\beta} a_i$, $b = \sum_{i \in S_\beta} b_i$, $b > 0$ and
$a > b \cdot 2^{\beta k/B(\tau, \rho)^2}$. We have that
\begin{displaymath}
D(\rho \| \sigma) = k \geq a \log \frac{a}{b}
                        >  \frac{\beta k a}{B(\tau, \rho)^2}
\Rightarrow
a < \frac{B(\tau, \rho)^2}{\beta}.
\end{displaymath}
Now, by the Cauchy-Schwarz inequality and the other inequalities
proved above, we get
\begin{eqnarray*}
B(\tau, \rho) & \leq & \sum_{i=1}^l \sqrt{c_i a_i} 
              \;  = \; \sum_{i \in S_\beta} \sqrt{c_i a_i} +
                       \sum_{i \not \in S_\beta} \sqrt{c_i a_i} \\
              & \leq & \sqrt{\sum_{i \in S_\beta} c_i}
                       \sqrt{\sum_{i \in S_\beta} a_i} +
                       2^{\beta k/(2 B(\tau, \rho)^2)}
                       \sum_{i \not \in S_\beta} \sqrt{c_i b_i} 
              \;\leq\; 1 \cdot \sqrt{a} + 2^{\beta k/(2 B(\tau, \rho)^2)}
                       B(\tau, \sigma) \\
              &   <  & \frac{B(\tau, \rho)}{\sqrt{\beta}} +
                       2^{\beta k/(2 B(\tau, \rho)^2)} B(\tau, \sigma).
\end{eqnarray*}
This shows that
\begin{displaymath}
B(\tau, \rho)^2 < (1 - \beta^{-1/2})^{-2} \cdot 
                  2^{\beta k/B(\tau, \rho)^2} B(\tau, \sigma)^2
\Rightarrow
|\braket{\psi}{\theta}|^2 < (1 - \beta^{-1/2})^{-2} \cdot 
                            2^{\beta k/|\braket{\psi}{\theta}|^2}
                            |\braket{\phi}{\theta}|^2.
\end{displaymath}
Since $k' = \beta k - 2 \log (1 - \beta^{-1/2})$, we get
$
|\braket{\phi}{\theta}|^2 \geq \frac{|\braket{\psi}{\theta}|^2}
                                    {2^{k'/|\braket{\psi}{\theta}|^2}},
$
completing the proof of the lemma.
\end{proof}

In the previous lemma, the purification $\ket{\phi}$ of $\sigma$
was a function of the POVM element $F$. We now prove a lemma which,
for any fixed $0 \leq p \leq 1$,
removes the dependence on $F$ satisfying $\Tr (F \ketbra{\psi}) \geq p$,
at the expense of having a, in general, mixed extension of $\sigma$ in the
place of a pure extension i.e. purification.
\begin{lemma}
\label{lem:lifting2}
Consider two Hilbert spaces $\cH$ and $\cK$,
$\dim(\cK) \geq \dim(\cH)$.
Let $\rho, \sigma$ be density matrices in $\cH$ and
$\ket{\psi}$ be a purification of $\rho$ in $\cH \otimes \cK$.
Let $0 \leq p \leq 1$ and $\beta > 1$. Then
there exists an extension $\omega$ of $\sigma$ in $\cH \otimes \cK$
such that for all
POVM elements $F$ on $\cH \otimes \cK$ such that
$\Tr (F \ketbra{\psi}) \geq p$,
$\Tr (F \omega) \geq p/2^{k'/p}$,
where $k' \defeq \beta D(\rho \| \sigma) - 2 \log (1 - \beta^{-1/2})$.
\end{lemma}
\begin{proof}
We assume without loss of
generality that $0 < D(\rho \| \sigma) < +\infty$ and that $p > 0$.
Consider the set $A_1$ of all extensions $\omega$ of $\sigma$ in
$\cH \otimes \cK$ and the set $A_2$ of all POVM operators $F$ in
$\cH \otimes \cK$ such that $\Tr (F \ketbra{\psi}) \geq p$. 
Observe that
$A_1$, $A_2$ are non-empty, compact, convex sets.
Without loss of generality, $A_2$ is non-empty.
The conditions of Fact~\ref{fact:minimax} are trivially satisfied
(note that we think
of our matrices, which in general have
complex entries, as vectors in a larger real vector space). Thus,
for every $F \in A_2$, we have a purification 
$\ket{\phi^F} \in \cH \otimes \cK$
of $\sigma$ such that
\begin{displaymath}
\Tr \left(F \ketbra{\phi^F}\right) \geq
       \frac{\Tr \left(F \ketbra{\psi}\right)}
            {2^{k'/\Tr \left(F \ketbra{\psi}\right)}} \geq
       \frac{p}{2^{k'/p}}.
\end{displaymath}
Using Fact~\ref{fact:minimax}, we see that there exists an extension
$\omega$ of $\sigma$ in $\cH \otimes \cK$ 
such that
$\Tr (F \omega) \geq \frac{p}{2^{k'/p}}$
for all $F \in A_1$. This completes the proof.
\end{proof}

The previous lemma depends upon the parameter $p$. We now remove
this restriction by performing a `discrete integration' operation and 
obtain an observational divergence `lifting' result, which may
be of independent interest.
\begin{lemma}[Observational divergence lifting]
\label{lem:lifting3}
Consider two Hilbert spaces $\cH, \cK$, 
$\dim(\cK) \geq \dim(\cH)$.
Let $\rho, \sigma$ be density matrices in $\cH$, and
$\ket{\psi}$ be a purification of $\rho$ in $\cH \otimes \cK$.
Then there exists an extension $\omega$ of $\sigma$ in $\cH \otimes \cK$
such that 
$
D(\left(\ketbra{\psi}\right) \| \omega) 
< D(\rho \| \sigma) + 4 \sqrt{D(\rho \| \sigma) + 1} + 
  2 \log (D(\rho \| \sigma) + 1) + 4.
$
\end{lemma}
\begin{proof}
We assume without loss of
generality that $0 < D(\rho \| \sigma) < +\infty$.
Let $\beta > 1$ and $\gamma \geq 1$.
Define the monotonically increasing
function $f: [0,1] \rightarrow [0,1]$ as follows:
\begin{displaymath}
f(p) \defeq \frac{p}{2^{k'/p}}
~~~ {\rm where} ~~~
0 \leq p \leq 1
~~~ {\rm and} ~~~
k' \defeq \beta D(\rho \| \sigma) - 2 \log (1 - \beta^{-1/2}).
\end{displaymath}
For a fixed positive integer $l$, define 
$T_\gamma(l) \defeq \sum_{i=1}^l l^{\gamma-1}$. It is easy to
see by elementary calculus that 
$
\gamma^{-1} \cdot l^\gamma 
\leq T_\gamma(l) 
\leq \gamma^{-1} \cdot (l+1)^\gamma.
$
Define
the density matrix $\omega_l$ in $\cH \otimes \cK$ as
$\omega_l \defeq (T_\gamma(l))^{-1} \sum_{i=1}^l i^{\gamma-1}\omega(i/l)$,
where for $0 \leq p \leq 1$, $\omega(p)$ is an extension of $\sigma$ in
$\cH \otimes \cK$ such that 
$\Tr (F \omega(p)) \geq f(p)$ for all POVM elements $F$ on
$\cH \otimes \cK$ satisfying $\Tr (F \ketbra{\psi}) \geq p$.
Such an $\omega(p)$ exists by Lemma~\ref{lem:lifting2}.
Then, $\parttr{\cK} \omega_l = \sigma$ i.e.
$\omega_l$ is an extension of $\sigma$ in $\cH \otimes \cK$.

Suppose $F$ is
a POVM element on $\cH \otimes \cK$. Let
$j/l \leq p \defeq \Tr (F \ketbra{\psi}) < (j+1)/l$, where
$0 \leq j \leq l$. We assume without loss of generality that
$p > 0$.  Then,
\begin{eqnarray*}
\Tr (F \omega_l) 
&   =  & \frac{1}{T_\gamma(l)}
         \sum_{i=1}^j i^{\gamma-1} \cdot \Tr (F \omega(i/l)) 
\;\geq\; \frac{1}{T_\gamma(l)} \sum_{i=1}^j i^{\gamma-1} \cdot f(i/l) \\
& \geq & \frac{T_\gamma(j)}{T_\gamma(l)} \cdot
         f\left(\frac{1}{T_\gamma(j)} \sum_{i=1}^j \frac{i^{\gamma}}{l}
          \right) 
\;  = \; \frac{T_\gamma(j)}{T_\gamma(l)} \cdot
         f\left(\frac{T_{\gamma+1}(j)}{l \cdot T_\gamma(j)}\right) \\
& \geq & \left(\frac{j}{l+1}\right)^\gamma \cdot 
         f\left(\frac{\gamma \cdot j^{\gamma+1}}
                     {l (\gamma+1) \cdot (j+1)^\gamma}
          \right) \\
& \geq & \left(\frac{pl-1}{l+1}\right)^\gamma \cdot 
         f\left(\left(\frac{\gamma (pl-1)}{(\gamma+1)l}\right) 
                \left(\frac{pl-1}{pl+1}\right)^\gamma
          \right).
\end{eqnarray*}
The second inequality above follows from the convexity of $f(\cdot)$.
By compactness, the set $\{\omega_l: l \in \naturals\}$ has limit points.
Choose a limit point point $\omega$. By standard continuity arguments,
$\parttr{\cK} \omega = \sigma$ and
\begin{eqnarray*}
q 
& \defeq & \Tr (F \omega) 
\; \geq \; \lim_{l \rightarrow +\infty}
           \left[
           \left(\frac{pl-1}{l+1}\right)^\gamma \cdot 
           f\left(\left(\frac{\gamma (pl-1)}{(\gamma+1)l}\right) 
                  \left(\frac{pl-1}{pl+1}\right)^\gamma
           \right)
           \right] 
\;   =  \; p^\gamma \cdot 
           f\left(\frac{\gamma p}{\gamma+1}\right) \\
&    =   & \frac{\gamma \cdot p^{\gamma+1}}
                {(\gamma+1) \cdot
                 2^{k'(\gamma+1) \gamma^{-1} p^{-1}}
                }.
\end{eqnarray*}
Hence, $q > 0$ and
\begin{eqnarray*}
p \log \frac{p}{q} 
& \leq & p \log 
         \left(\gamma^{-1} (\gamma+1) \cdot p^{-\gamma} \cdot 
               2^{k'(\gamma+1) \gamma^{-1} p^{-1}}
         \right) 
\;  = \; p \log (1 + \gamma^{-1}) - \gamma p \log p +
         (1 + \gamma^{-1}) k' \\
&   <  & (1 + \gamma^{-1}) k' + \gamma + 1.
\end{eqnarray*}
The second inequality follows because $- p \log p < 1$ for
$0 \leq p \leq 1$, and $\log (1 + \gamma^{-1}) \leq 1$ for all
$\gamma \geq 1$. Substituting 
$k' = \beta D(\rho \| \sigma) - 2 \log (1 - \beta^{-1/2})$ gives
\[
D(\left(\ketbra{\psi}\right) \| \omega) 
< \beta (1 + \gamma^{-1}) D(\rho \| \sigma) - 
  2 (1 + \gamma^{-1}) \log (1 - \beta^{-1/2}) + \gamma + 1.
\]
We set $\beta = (1 + (D(\rho \| \sigma)+1)^{-1/2})^2$ and
$\gamma = (D(\rho \| \sigma)+1)^{1/2}$ to get
\begin{eqnarray*}
D(\left(\ketbra{\psi}\right) \| \omega) 
& < & (1 + (D(\rho \| \sigma)+1)^{-1/2})^2 \cdot
      (1 + (D(\rho \| \sigma)+1)^{-1/2}) \cdot D(\rho \| \sigma) \\
&   & {} +
      (1 + (D(\rho \| \sigma)+1)^{-1/2}) \cdot
      \log (D(\rho \| \sigma)+1) + (D(\rho \| \sigma)+1)^{1/2} + 1 \\
& < & D(\rho \| \sigma) + 4 \sqrt{D(\rho \| \sigma)+1} +
      (1 + (D(\rho \| \sigma)+1)^{-1/2}) \cdot
      \log (D(\rho \| \sigma)+1) + 4 \\
& < & D(\rho \| \sigma) + 4 \sqrt{D(\rho \| \sigma)+1} +
      2 \log (D(\rho \| \sigma)+1) + 4.
\end{eqnarray*}
This completes the proof of the lemma.
\end{proof}

Lemma~\ref{lem:lifting3} relates the observational
divergence of a pair of density matrices
to the observational 
divergence of their extensions in an extended Hilbert space,
where the extension
of the first density matrix is a pure state. Using this,
we are now finally in a position to prove 
the quantum substate theorem.  

\bigskip

\noindent{\bf Proof (Theorem~\ref{thm:substate}):}
By Proposition~\ref{prop:divrelentropy} and Lemma~\ref{lem:lifting3}, 
there exists a density matrix $\omega$ in $\cH \otimes \cK$ such that
$\parttr{\cK} \omega = \sigma$ and
\begin{eqnarray*}
D\left(\left(\ketbra{\psi}\right) \| \, \omega\right)
& < & D(\rho \| \sigma) + 4 \sqrt{D(\rho \| \sigma)+1} + 
      2 \log (D(\rho \| \sigma)+1) + 4 \\
& < & S(\rho \| \sigma) + 4 \sqrt{S(\rho \| \sigma) + 2} +
      2 \log (S(\rho \| \sigma)+2) + 5
\;=\;k'.
\end{eqnarray*}
By Lemma~\ref{lem:twobytwo}, there exists a pure state $\ket{\phi}$
such that
\begin{displaymath}
\trnorm{\ketbra{\psi} - \ketbra{\phi}} \leq \frac{2}{\sqrt{r}}
~~~ {\rm and} ~~~
\left(\frac{r-1}{r 2^{rk'}}\right) \ketbra{\phi} \leq \omega.
\end{displaymath}
Let $\tau_1 \defeq \parttr{\cK} \ketbra{\phi}$. By above, 
$
\left(\frac{r-1}{r 2^{rk'}}\right) \tau_1 \leq \sigma.
$
That is,
there exists a density matrix $\tau_2$ in $\cH$ such that
\begin{displaymath}
\sigma = \left( \frac{r-1}{r2^{rk'}} \right) \tau_1 +
         \left( 1 - \frac{r-1}{r2^{rk'}} \right) \tau_2.
\end{displaymath}
Let $\ket{\theta} \in \cH \otimes \cK$ be a canonical purification
of $\tau_2$. Then, $\ket{\zeta}$ defined in the statement of 
Theorem~\ref{thm:substate} is a purification
of $\sigma$ in $\cH \otimes \cK \otimes \complexes^2$.
This completes the proof of Theorem~\ref{thm:substate}.
\qed

\section{Conclusion and open problems}
\label{sec:conclusion}
In this paper we have proved a theorem about relative entropy of
quantum states which gives a novel interpretation to this
information theoretic quantity. Using this theorem, we have shown a
privacy trade-off for computing set membership in the two-party
quantum communication model. 

The statements of the classical and quantum substate theorems have
one important difference. For two quantum states $\rho$, $\sigma$
with $S(\rho \| \sigma) = k$, the distance between $\rho$ and $\rho'$,
where $\rho'/2^{O(k)} \leq \sigma$, is less in the classical case
than in the quantum case. More formally,
the dependence on $r$ in Theorem~\ref{thm:substate} is 
$O(1/\sqrt{r})$ whereas in the classical analogue,
Result~\ref{res:substate}', the dependence is like $O(1/r)$. The better
dependence in the classical scenario enables us to prove a kind
of converse to the classical substate theorem, which is outlined in
the appendix. It will
be interesting to see if the dependence in the quantum setting
can be improved to match the classical case, enabling us to prove a
similar quantum converse. 

Another open question is if there is an alternate
proof for the quantum substate theorem which
does not go through observational divergence lifting.
Finally, it will also be interesting to see find yet more applications
of the classical and quantum substate theorems.

\subsection*{Acknowledgements}
We are very grateful to Ashwin Nayak for his contribution to this
work. He patiently went through several versions of our proofs; his
counter examples and insights were invaluable in arriving at our
definition of privacy. We are also grateful to K.~R.~Parthasarathy and
Rajendra Bhatia for sharing with us their insights in 
operator theory.

\newcommand{\etalchar}[1]{$^{#1}$}

\appendix 

\section{Relationships between three distinguishability measures}
In this paper we have seen two measures of
distinguishability between quantum states viz. relative entropy and
observational divergence. The substate theorem gives a connection between
observational divergence and a third measure of distinguishability
between quantum states, which we call the {\em substate property}.
We define three variants of the substate property below, and study
the relationships between them and relative entropy and observational
divergence.
\begin{definition}[Substate property]
Let $\rho$, $\sigma$ be two quantum states in the same Hilbert space $\cH$.
They are said to have the {\em $k$-substate property}
if for all $r \geq 1$, there exists a quantum state $\rho(r)$ in
$\cH$ such that $\trnorm{\rho - \rho(r)} \leq 2/r$ and
$\left(\frac{r-1}{r 2^{rk}}\right) \rho(r) \leq \sigma$. They
are said to have the {\em weak} $k$-substate property if
$\trnorm{\rho - \rho(r)}$ is upper bounded by $2/\sqrt{r}$ instead
of $2/r$. They are said to have the {\em strong} $k$-substate property
if $\rho/2^k \leq \sigma$.
\end{definition}
The next proposition lists some easy consequences of the definition
of substate property.
\begin{proposition}
\label{prop:propsubstate}
Let $\rho, \sigma$ be density matrices in the same 
Hilbert space $\cH$. Then
\begin{enumerate}
\item If $\rho, \sigma$ satisfy the $k$-substate property, then $k \geq 0$ 
      with equality iff $\rho = \sigma$.
\item $\rho, \sigma$ satisfy the $k$-substate property with $k < +\infty$
      iff $\supp(\rho) \subseteq \supp(\sigma)$.
\item {\em (Unitary invariance)} If $U$ is a unitary transformation on
      $\cH$, then $\rho, \sigma$ satisfy the $k$-substate property
      iff $U \rho, U \sigma$ satisfy the $k$-substate property.
\item {\em (Monotonicity)} Suppose $\cK$ is a Hilbert
      space, and $\rho', \sigma'$ are extensions of $\rho, \sigma$ in
      $\cH \otimes \cK$. If $\rho', \sigma'$ satisfy the $k$-substate
      property, then $\rho, \sigma$ satisfy it also. This implies,
      via unitary invariance and the Kraus representation theorem,
      that if $\cT$ is a completely positive trace preserving
      superoperator from $\cH$ to a Hilbert space $\cL$, then
      if $\rho, \sigma$ satisfy the $k$-substate
      property, $\cT \rho, \cT \sigma$ do so also.
\end{enumerate}
Similar statements hold for the weak and strong $k$-substate property
also.
\end{proposition}

The following proposition states various relationships between our three
measures of distinguishability that we have mentioned earlier.
\begin{proposition}
\label{prop:divrelsub}
We have:
\begin{enumerate}

\item
\label{item:classicalsubstate}
{\em (Classical substate theorem)}
Two probability distributions $P, Q$ on $[n]$ with $D(P \| Q) = k$
satisfy the $k$-substate property.

\item 
\label{item:quantumsubstate}
{\em (Quantum substate theorem)}
Two quantum states $\rho, \sigma$ in $\complexes^n$ with 
$D(\rho \| \sigma) = k$ satisfy the weak $k'$-substate property with 
$k' = k + 4\sqrt{k+1} + 2\log (k+1) + 4$.

\item 
\label{item:substateconverse}
If quantum states $\rho, \sigma$ in $\complexes^n$ have the 
$k$-substate property, then $D(\rho \| \sigma) \leq 2k +2$.

\item 
\label{item:substateweakconverse}
If quantum states $\rho, \sigma$ in $\complexes^n$ have the strong 
$k$-substate property, then $S(\rho \| \sigma) \leq k$.

\item 
\label{item:upperboundSclassical}
For any probability distributions $P, Q$ on $[n]$, 
$D(P \| Q) - 1 \leq S(P \| Q) \leq D(P \| Q) (n - 1)$.

\item 
\label{item:upperboundSquantum}
For any quantum states $\rho, \sigma$ in $\complexes^n$, 
$
D(\rho \| \sigma) - 1 
\leq S(\rho \| \sigma) 
\leq   D(\rho \| \sigma) (n - 1) + \log n$.

\item
\label{item:lowerboundSclassical}
There exist probability distributions $P, Q$ on $[n]$ such that
$S(P \| Q) > \left(\frac{D(P \| Q)}{2} - 1\right) (n -2) - 1$.

\item 
\label{item:monotonicitySquantum}
For any two quantum states $\rho, \sigma$ in $\complexes^n$, there exists
a two-outcome POVM $\cF$ on $\complexes^n$ such that
$
S(\rho\|\sigma) 
\geq S(\cF\rho\|\cF\sigma) 
\geq \frac{S(\rho\|\sigma)-\log n}{n-1} - 1.
$
\end{enumerate}
\end{proposition}

\noindent {\bf Remarks:} \\
1.\ From Parts~\ref{item:classicalsubstate} and 
\ref{item:substateweakconverse} of Proposition~\ref{prop:divrelsub}, we 
see that the classical substate theorem (Result~\ref{res:substate}') 
has a converse. \\
2.\ Unfortunately, we are
unable to prove a converse to the quantum substate theorem 
(Result~\ref{res:substate}) as 
Part~\ref{item:quantumsubstate} of Proposition~\ref{prop:divrelsub}
only guarantees a weak substate property between the two
quantum states $\rho, \sigma$. \\
3.\ Part~\ref{item:monotonicitySquantum} of 
Proposition~\ref{prop:divrelsub} is a counterpart to monotonicity
of relative entropy (Fact~\ref{fact:relentprop}).

\bigskip

\noindent{\bf Proof (Proposition~\ref{prop:divrelsub}):}

\begin{enumerate}

\item
Without loss of generality, $k > 0$. Let $r \geq 1$.
Define the set 
${\sf Bad} \defeq \{i \in [n]: P(i)/2^{rk} > Q(i)\}$. Then,
\[
k 
  =  D(P \| Q) 
\geq P({\sf Bad}) \log \frac{P({\sf Bad})}{Q({\sf Bad})}
  >  P({\sf Bad}) \cdot rk
\Rightarrow
P({\sf Bad}) < \frac{1}{r},
\]
which is the same as expression~(\ref{eq:weaksubstate}) in
Section~\ref{subsec:introsubstate}. We can now
argue similarly as in the proof of Result~\ref{res:substate}' to
prove Part~\ref{item:classicalsubstate} of the present proposition.

\item
Follows from Lemmas~\ref{lem:lifting3} and \ref{lem:twobytwo}.

\item 
Without loss of generality, 
$0 < k_1 \defeq D(\rho \| \sigma) < +\infty$.
Let $F$ be a POVM element in $\complexes^n$ such that 
\[
k_1 = p \log (p/q)
\Rightarrow 
q = \frac{p}{2^{k_1/p}},
\]
where
$p \defeq \Tr (F \rho)$ and $ q \defeq \Tr (F \sigma)$. Note that
$p > 0$.
Let $r \defeq 2/p$. Since $\rho, \sigma$ have the $k$-substate
property, let $\rho'$ be the quantum state in $\complexes^n$ such that
$\trnorm{\rho-\rho'} \leq \frac{2}{r} = p$
and 
$\left(\frac{r-1}{r 2^{rk}}\right) \rho' \leq \sigma$.
Define $p' \defeq \Tr (F \rho')$. Then, $p' \geq p/2$.
Also, 
\begin{eqnarray*}
\frac{p}{2^{k_1/p}}
&   =  & q  
\;  = \; \Tr (F \sigma) 
\;\geq\; \left(\frac{1-r^{-1}}{2^{rk}}\right) \Tr (F \rho') 
\;  = \; \left(1-\frac{p}{2}\right) \frac{p'}{2^{rk}} \\
& \geq & \frac{p}{2^{rk + 2}}.
\end{eqnarray*}
The last inequality above follows because $p \leq 1$ and $p' \geq p/2$.
This implies that
\[
rk+2 \geq \frac{k_1}{p} 
\Rightarrow p(rk+2) \geq k_1 
\Rightarrow 2k+ 2 \geq k_1, 
\]
where the second implication follows because $p \leq 1$ and $p = 2/r$.
This completes the proof of Part~\ref{item:substateconverse}
of the present proposition.

\item
Without loss of generality, $k < +\infty$.
We have
\[
S(\rho \| \sigma) 
  =  \Tr \rho \log \rho - \Tr \rho \log \sigma 
\leq \Tr \rho \log \rho - \Tr \rho \log \frac{\rho}{2^{k}}
  =  k \cdot \Tr \rho = k.
\]
The inequality above is by monotonicity of the
logarithm function on positive operators~\cite{lowner:logmonotone}.

\item
Without loss of generality, $0 < D(P \| Q) < +\infty$.
The lower bound on $S(P \| Q)$ was proved in 
Proposition~\ref{prop:divrelentropy}.
Define $x_i = \log (p_i/q_i)$. 
We can assume without loss of generality, by perturbing $Q$ slightly,
that the values $x_i$ are distinct for distinct $i$.
Let $S' = \{i : x_i > 0\}$. 
Let $k \defeq D(P \| Q)$. Let 
For all positive $l$, define 
$S_l \defeq \{ i \in [n] : x_i \geq l\}$.
Therefore,
\[
k \geq \Pr_P[S_l] \log \frac{\Pr_P[S_l]}{\Pr_Q[S_l]} \geq \Pr_P[S_l] l
\Rightarrow 
\Pr_P[S_l] \leq  k/l.
\]
Assume without loss of generality that $x_1 < x_2 < \cdots < x_n$.
Then if $x_i > 0$, $\Pr_P[S_{x_i}] \leq k/x_i$.
Since $S(P \| Q) \leq \sum_{i \in S'} p_i x_i$, the upper bound on 
$S(P \| Q)$
is maximised when $S' = \{2, \ldots, n\}$, $p_n = k/x_n$, 
$p_i = k (1/x_i - 1/x_{i+1})$ for all $i \in \{2, \ldots, n - 1\}$,
and $p_1 = 1 - \sum_{i=2}^n p_i$. Then,
\begin{eqnarray*}
S(P \| Q) 
& \leq & \sum_{i=2}^{n} p_i x_i 
\;  = \; k \sum_{i=2}^{n-1} x_i (1/x_i - 1/x_{i+1}) + k
\;  = \; k \sum_{i=2}^{n-1} \frac{x_{i+1} - x_i}{x_{i+1}} + k
\;\leq\; k \sum_{i=2}^{n-1} 1 + k \\
&   =  & k (n - 1).
\end{eqnarray*}

\item 
Without loss of generality, $0 < D(\rho \| \sigma) < +\infty$.
The lower bound on $S(\rho \| \sigma)$ was proved in 
Proposition~\ref{prop:divrelentropy}.
Let us measure $\rho$ and $\sigma$ in the eigenbasis of
$\sigma$. We get two distributions, $P$ and $Q$. Below, we will
sometimes think of $P, Q$ as diagonal density matrices.
From Part~\ref{item:upperboundSclassical}
of the present proposition, it follows that
\begin{eqnarray*}
D(P \| Q) (n - 1) 
& \geq & S(P | Q) 
\;  = \; \Tr (P \log P) - \Tr (P \log Q) 
\;\geq\; - \log n - \Tr (P \log Q) \\
&   =  &  - \log n - \Tr (\rho \log \sigma) 
\;  = \;  - \log n + S(\rho \| \sigma) - \Tr (\rho \log \rho) \\
& \geq &  - \log n + S(\rho \| \sigma).
\end{eqnarray*}
The second equality above holds since the measurement was in the
eigenbasis of $\sigma$.

Thus, 
\[
S(\rho \| \sigma) 
\leq D(P \| Q) (n - 1) + \log n 
\leq D(\rho \| \sigma) (n - 1) + \log n,
\] 
where the second inequality is by monotonicity of observational
divergence (Proposition~\ref{prop:propdiv}).

\item
Fix $a > 1$, $k > 0$.
Define for all $i \in \{2, \ldots, n-1\}$, $p_i \defeq a^{-i} (a - 1)$,
and $p_1 \defeq a^{-1} (a - 1)$, $p_n \defeq a^{-(n-1)}$.
Define
for all $i \in \{2, \ldots, n\}$, $q_i \defeq p_i 2^{-k a^{i-1}}$,
and $q_1 \defeq 1 - \sum_{i=2}^n q_i$. Define
$P \defeq (p_1, \ldots, p_n)$, $Q \defeq (q_1, \ldots, q_n)$; $P, Q$
are probability distributions on $[n]$.
For any $r > 1$, consider 
$
\tilde{P}\defeq (p_1, \ldots, p_{\lceil \log_a r\rceil+1}, 0, \ldots, 0)$ 
normalised to make it a probability distribution on $[n]$. It is easy 
to see that $\totvar{P - \tilde{P}} \leq 2/r$ and 
$\frac{(r-1) \tilde{P}}{r 2^{rk}} \leq Q$. This shows that $P, Q$ 
satisfy the $k$-substate property, hence 
$D(P \| Q) \leq 2(k+1)$ by Part~\ref{item:substateconverse}
of the present proposition.
 
Now,
\begin{eqnarray*}
S( P \| Q) 
&   =  & \sum_{i=1}^n p_i \log \frac{p_i}{q_i} 
\;\geq\; p_1 \log p_1 + \sum_{i=2}^n p_i \log \frac{p_i}{q_i} 
\;  > \; -1 + (n-2) \frac{k(a-1)}{a} + k \\
&   =  & k (n - 1) - \frac{k(n-2)}{a} - 1.
\end{eqnarray*}
The second inequality above follows because $p \log p > -1$ for all
$0 \leq p \leq 1$.
By choosing $a$ large enough, we can achieve 
$S(P \| Q) > k (n - 2) - 1$. This completes the proof of
Part~\ref{item:lowerboundSclassical} of the present proposition.

\item
The upper bound on $S(\cF \rho \| \cF \sigma)$ follows from
the monotonicity of relative entropy (Fact~\ref{fact:relentprop}).
Without loss of generality, $0 < S(\rho \| \sigma) < +\infty$.
We know that there exists a POVM element $F$ in $\complexes^n$ such that
$D(\rho \| \sigma) = p \log (p/q)$, where
$p \defeq \Tr F \rho$ and $q \defeq \Tr F \sigma$. 
Define the two-outcome POVM $\cF$ on $\complexes^n$ to be
$(F, \identity - F)$, where $\identity$ is the identity operator on
$\complexes^n$.
Then, the probability distributions $\cF \rho = (p, 1-p)$ and 
$\cF \sigma = (q, 1-q)$. 
Note that 
\[
S(\cF \rho \| \cF \sigma) 
= p \log \frac{p}{q} + (1-p) \log \frac{1-p}{1-q} 
> p \log \frac{p}{q}  - 1 \\
= D(\rho \| \sigma)  - 1,
\]
where the inequality follows because $x \log x > -1$ for all
$0 \leq x \leq 1$.
From Part~\ref{item:upperboundSquantum} of the present proposition,
it follows that 
\begin{eqnarray*}
S(\rho \| \sigma) 
& \leq & D(\rho \| \sigma) (n - 1) + \log n 
\;\leq\; (S(\cF \rho \| \cF \sigma) + 1) (n - 1) + \log n \\
\Rightarrow  S(\cF \rho \| \cF \sigma) 
& \geq & \frac{S(\rho \| \sigma) - \log n}{n-1} - 1.
\end{eqnarray*}

\end{enumerate}

\qed

\end{document}